\documentclass[a4paper,12pt]{article}

\usepackage{amsmath}
\usepackage[psamsfonts]{amssymb}
\usepackage{rsfs}
\usepackage{bm}

\usepackage{cite}
\usepackage[dvips]{graphicx}
\usepackage{color}

\makeatletter
\@addtoreset{equation}{section}
\makeatother

\begin{document} 

\begin{titlepage}

\baselineskip 10pt
\hrule 
\vskip 5pt
\leftline{}
\leftline{Chiba Univ./KEK Preprint
          \hfill   \small \hbox{\bf CHIBA-EP-172}}
\leftline{\hfill   \small \hbox{\bf KEK Preprint 2008-14}}
\leftline{\hfill   \small \hbox{August 2008}}
\vskip 5pt
\baselineskip 14pt
\hrule 
\vskip 1.0cm
\centerline{\Large\bf 
} 
\vskip 0.3cm
\centerline{\Large\bf  
Magnetic monopole loops  
}
\vskip 0.3cm
\centerline{\Large\bf  
supported by a  meron pair 
}
\vskip 0.3cm
\centerline{\Large\bf  
as the quark confiner
}

\vskip 1.0cm

\centerline{{\bf 
Kei-Ichi Kondo$^{\dagger,{1}}$,   
Nobuyuki Fukui$^{\dagger,{2}}$,  
Akihiro Shibata$^{\flat,{3}}$  
and
Toru Shinohara$^{\dagger,{4}}$, 
}}  
\vskip 0.5cm
\centerline{\it
${}^{\dagger}$Department of Physics, Graduate School of Science, 
}
\centerline{\it
Chiba University, Chiba 263-8522, Japan
}
\vskip 0.3cm
\centerline{\it
${}^{\flat}$Computing Research Center, High Energy Accelerator Research Organization (KEK)   
}
\vskip 0.1cm
\centerline{\it
\& 
Graduate Univ. for Advanced Studies (Sokendai), 
Tsukuba 305-0801, 
Japan
}

\vskip 1cm

\begin{abstract}
We give an analytical solution representing circular magnetic monopole loops joining a pair of merons in the four-dimensional Euclidean SU(2) Yang-Mills theory. 
This is achieved by solving the differential equation for the adjoint color (magnetic monopole) field  in the two--meron background field  within the recently developed reformulation of the Yang-Mills theory. 
Our analytical solution corresponds to the numerical solution found by Montero and Negele on a lattice. 
This result strongly suggests that a meron pair is the most relevant quark confiner in the original Yang-Mills theory, as Callan, Dashen and Gross suggested long ago. 
 
\end{abstract}

 
\vskip 0.5cm

\hrule  
\vskip 0.1cm
${}^1$ 
  E-mail:  {\tt kondok@faculty.chiba-u.jp}
  
${}^2$ 
  E-mail:  {\tt  n.fukui@graduate.chiba-u.jp}
  
${}^3$ 
  E-mail:  {\tt akihiro.shibata@kek.jp} 

${}^4$ 
  E-mail:  {\tt sinohara@graduate.chiba-u.jp}

\par 
\par\noindent


\vskip 0.5cm

\newpage
\pagenumbering{roman}




\end{titlepage}


\pagenumbering{arabic}

\baselineskip 14pt

\section{Introduction}

The dual superconductivity picture \cite{dualsuper} for quark confinement  \cite{Wilson74} was proposed long ago and it is now believed to be a promising mechanism for quark confinement. 
The dual superconductivity is supposed to be realized as an electric-magnetic dual of the ordinary superconductivity.  For this to be possible, there must exist magnetic monopoles to be condensed for causing the dual Meissner effect, just as the  Cooper pairs exist and they are condensed to cause Meissner effect in the ordinary superconductivity. 
The idea of dual superconductivity is intuitively easy to understand, but upgrading this idea into a quantitative theory  was not so easy, as can be seen from a fact that we are still involved in this work.

The possible topological soliton in pure Yang-Mills theory \cite{YM54} (with no matter fields) is only the Yang-Mills instanton \cite{BPST,CF77,tHooft76,Wilczek77} with a finite action integral and integer Pontryagin index in $D=4$ dimensional Euclidean spacetime where the continuous map $U$: $S^3 \rightarrow SU(2) \simeq S^3$ is classified by the Homotopy  $\pi_3(S^3)=\mathbb{Z}$.
\footnote{
The Yang-Mills instanton can be also regarded as a solution with a finite energy in D=4+1 Minkowski spacetime.
Here we use a topological soliton as implying a solution of the Yang-Mills field  equation (of motion) having an invariant under the continuous deformation of the solutions, e.g.,  characterized by non-trivial Homotopy class. 
}
However, we have some arguments suggesting that Yang-Mills  instantons do not confine quarks in four dimensions, e.g.,  \cite{Witten79}. 
This is in sharp contrast to the $D=3$ case. For instance, in the Georgi-Glashow model, point-like magnetic monopoles exist as instantons in three dimensions, leading to the area law of the Wilson loop average \cite{Polyakov77}.

In view of this, 't Hooft \cite{tHooft81} has proposed an explicit prescription which enables one to extract (Abelian) magnetic monopoles from the Yang-Mills theory as gauge-fixing defects, which is called the Abelian projection method. 
In this prescription,  the location of an Abelian magnetic monopole in $SU(2)$ Yang-Mills theory is specified by the simultaneous zeros (of first order) in $\mathbb{R}^D$ of a three-component field $\vec{\phi}(x)=\{ \phi^A(x) \}_{A=1,2,3}$, which we call the monopole field hereafter.  
As a result, an Abelian magnetic monopole is a topological object of co-dimension 3 if they exist at all, characterized by a continuous map $S^2 \rightarrow SU(2)/U(1) \simeq S^2$ with $\pi_2(S^2)=\mathbb{Z}$. Therefore, an Abelian magnetic monopole is represented by a point in $D=3$, as expected.
In $D=4$ dimensions the world line of a magnetic monopole must draw a closed loop, not merely an open line, due to the topological conservation law of the magnetic monopole current. (A magnetic current  extending from an infinity to another infinity can be also identified with a closed loop.) 
It is known that large magnetic monopole loops are the most dominant configurations responsible for confinement. See e.g., \cite{CP97} for a review. 

The second obstacle is the lack of information as to how the magnetic monopole is related to the original Yang-Mills field in a gauge-invariant manner. 
In the 't Hooft proposal, the monopole field $\vec{\phi}(x)$ is an arbitrary composite operator of the Yang-Mills field $\mathscr{A}^A_\mu(x)$ as long as it transforms according to the adjoint representation under the gauge transformation, e.g., $\phi^A=\mathscr{F}_{12}^A$ (a component of the field strength $\mathscr{F}_{\mu\nu}^A$).  
The maximal Abelian gauge (MAG) \cite{KLSW87} is well-known to be the most effective choice in practical calculations,  corresponding to a specific choice for the monopole field in  the Abelian projection method. 
The MAG is given by minimizing the gauge-fixing functional written in terms of the off-diagonal gluon fields $A_\mu^a$ ($a=1,2$):
$
 F_{\rm MAG} = \int d^4 \frac12 [(A_\mu^1)^2+(A_\mu^2)^2]
$ 
using the gauge degrees of freedom to transform the gauge field variable as close as possible to the maximal torus group $U(1)$.

It is important to establish the relationship between the Abelian magnetic monopole in question and an original Yang-Mills field.  For this purpose, some works have already been devoted to constructing  explicit configurations of the magnetic monopole loops from  appropriately chosen configurations of the Yang-Mills field.  
In fact, Chernodub and Gubarev \cite{CG95} have pointed that one instanton or the set of instantons arranged along a straight line induce an Abelian  magnetic monopole current along a straight line going through centers of instantons within  MAG. 
In this case, the magnetic monopole is given by the standard static hedgehog configuration. 
However, this solution yields a divergent value for the gauge-fixing function of MAG:
$
 F_{\rm MAG} = \int d^4 \frac12 [(A_\mu^1)^2+(A_\mu^2)^2]
$. 
Therefore, it must be excluded in four dimensions.

A laborious and important work has been done by Brower, Orginos and Tan (BOT) \cite{BOT97} who investigated within the MAG whether the magnetic monopole loop represented by a circle with a non-zero and finite radius can exist for some given instanton configurations or not. 
They concluded the absence of such a stable magnetic monopole loop for one-instanton Yang-Mills background: a circular magnetic monopole loop centered on an instanton is inevitably shrank to the center point, if one imposes the condition of minimizing the MAG functional
$
 F_{\rm MAG} 
$. 
While an instanton-antiinstanton pair seems to support a stable magnetic monopole loop, although it is not a solution of Yang-Mills equation of motion.  
These results suggest that instantons are not the topological objects responsible for quark confinement from the viewpoint of the dual superconductivity. 
Moreover, these conclusions heavily rely on their hard work of numerically solving partial differential equations. 
See e.g., \cite{HT96} for the corresponding result obtained from numerical simulations on a lattice. 

Subsequently, Bruckmann, Heinzl, Vekua and Wipf (BHVW) \cite{BHVW01} have performed a systematic and analytical treatment to this problem within the Laplacian Abelian gauge (LAG) \cite{LAG}, which has succeeded to shed new light on the relationship between an Abelian magnetic monopole and an instanton from a different angle.
In the LAG, zeros of an auxiliary Higgs field $\vec{\phi}(x)$ induce topological defects in the gauge potential, upon diagonalization. 
It is important to notice that the nature of the defects depends on the order of the zeros.  For first-order zeros, one obtains magnetic monopoles. The defects from zeros of second order are Hopfion which is characterized by a topological invariant called Hopf index \cite{Hopf31} for the Hopf map $S^3 \rightarrow SU(2)/U(1) \simeq S^2$ with non-trivial Homotopy $\pi_3(S^2)=\mathbb{Z}$.
They have solved the eigenvalue problem of the covariant Laplacian $-D_\mu[\mathbf{A}]D_\mu[\mathbf{A}]$ in the adjoint representation in the background of a single instanton $\mathbf{A}$ (in the singular gauge):
$
 -D_\mu[\mathbf{A}]D_\mu[\mathbf{A}]\vec{\phi}(x)=\lambda \vec{\phi}(x)
$
 and have obtained the auxiliary Higgs field $\vec{\phi}(x)$ as the (normalizable) ground state wave function having the lowest eigenvalue $\lambda$.  
Consequently, they have found that the auxiliary Higgs field $\vec{\phi}(x)$ is given by the standard Hopf map: $S^3 \rightarrow S^2$ in the neighborhood of the center of an instanton and by a constant, e.g., $(0,0,1)$ after normalization in the distant region far away from the center where two regions are separated by the scale of the instanton size parameter $\rho$. 
The zeros of the auxiliary Higgs field $\vec{\phi}(x)$ agree with the origin.  This results enable one to explain the BOT result without numerical calculations.  
In BHVW, however,  $\mathbb{R}^4$ was replaced by a four sphere $S^4$ of a finite radius in order to obtain a finite LAG functional (convergent integral). 
See also \cite{Reinhardt97,Jahn00,TTF00,HY04,BH03} for relationships among various topological objects.

In the course of studying the relationship between Abelian magnetic monopole and center vortices \cite{Greensite03}, merons \cite{AFF76} are recognized as important object \cite{ER00,Cornwall98}.
Merons \cite{AFF76,AFF77} are solutions of the Yang-Mills field equation and are characterized by one half topological charge, i.e., having half-integer Pontryagin index. 
These configurations escaped from the above consideration, since they have  infinite action due to their singular behaviors.  However, once they receive an ultraviolet regularization which does not influence quark confinement, they can have finite action and contribute to the functional integration over the Yang-Mills field in calculating the Wilson loop average, based on action and entropy argument in the strong coupling region above a critical value. 
In fact, Callan, Dashen and Gross \cite{CDG78} have discussed that merons are the most dominant quark confiner. 
In fact, Reinhardt and Tok \cite{RT01} have investigated the relationship among Abelian magnetic monopoles and center vortices in various Yang-Mills back ground fields: one meron, one instanton, instanton-antiinstanton pair, using both the LAG and the Laplacian center gauge (LCG).
It has been pointed out that  Abelian magnetic monopole and meron pair are mediated by sheets of center vortices \cite{ER00}.
In fact, Montero and Negele \cite{MN02} have obtained an Abelian magnetic monopole loop and center vortices for two merons (meron pair) by using numerical simulations on a lattice, see also \cite{meron} for related works. 

The third obstacle in these approaches lies in a fact that topological objects such as Abelian magnetic monopoles and center vortices are obtained as gauge fixing defects. 
Therefore, they are not free from criticism of gauge artifacts.%
\footnote{
There is an approach to gauge-invariant Abelian confinement mechanism, see \cite{Suzuki08}. 
}
Recently, we have given a gauge-invariant (gauge independent) definition of magnetic monopoles  \cite{KMS06,KSM08,Kondo08} and vortices \cite{Kondo08b} in Yang-Mills theory in the framework of a new reformulation of Yang-Mills theory based on change of field variables  
 founded in \cite{Cho80}.  
The lattice version has been constructed to support them by numerical simulations \cite{KKMSSI05}. 
These are suggested from a non-Abelian Stokes theorem for the Wilson loop operator \cite{DP89,KondoIV,Kondo08}. 

This paper is organized as follows. 
In sections 2 and 3, we investigate how the gauge-invariant magnetic monopoles are obtained analytically for a given Yang-Mills background field in four dimensions.
In sections 4 and 5, we show that the previous results \cite{BOT97,BHVW01,RT01} obtained for one-meron and one-instanton  are easily reproduced within our reformulation. 
In sections 6, we give a new result for  gauge-invariant magnetic monopole loops in Yang-Mills theory:
It is shown in an analytical way that  there exist circular magnetic monopole loops joining two merons. 
This will be the first analytical solution of stable magnetic monopole loops constructed from the Yang-Mills field with non-trivial but finite Pontryagin index.
(Bruckmann and Hansen  \cite{BH03} constructed a ring of magnetic monopole by superposing  infinitely many instantons on a circle. This configuration has infinite action and infinite Pontrayagin index.) 
The analytical solution for  magnetic monopole loops given in this paper correspond to the numerical solution found by Montero and Negele \cite{MN02} on a lattice.

The result in this paper has rather interesting implications to  the quark confinement mechanism. As mentioned above, the gauge-invariant magnetic monopole is a complicated object obtained by the non-linear change of variables from the original Yang-Mills field, although they are fundamental objects necessary for the naive dual-superconductivity scenario of quark confinement. 
In other words, the result in this paper indicates that a meron pair is the most relevant quark confiner if viewed from the original Yang-Mills theory, as Callan, Dashen and Gross \cite{CDG78} suggested long ago.

\section{Reduction condition}

In a previous paper \cite{KMS06}, we have given a prescription for obtaining the gauge-invariant magnetic monopole from the original Yang-Mills field $\mathbf{A}_\mu(x)$. 

(i) For a given SU(2) Yang-Mills field $\mathbf{A}_\mu(x)=\mathbf{A}^A_\mu(x)\frac{\sigma_A}{2}$,  the color field $\mathbf{n}(x)$ is obtained by solving the differential equation:
\begin{align}
 \mathbf{n}(x) \times D_\mu[\mathbf{A}]D_\mu[\mathbf{A}] \mathbf{n}(x) = \mathbf{0} 
 ,
 \label{RDE1}
\end{align}
which we call the reduction differential equation (RDE).
Here the color field has the unit length
\begin{align}
   \mathbf{n}(x) \cdot \mathbf{n}(x) = 1
   .
   \label{unit}
\end{align}

(ii) Once  the color field $\mathbf{n}(x)$ is known, the gauge-invariant ``magnetic-monopole current''   $k$  is constructed  by applying the exterior derivative $d$, the coderivative (adjoint derivative) $\delta$ and Hodge star operation $*$ to $f$:
\begin{align}
 k:=  \delta *f = *df 
  ,
  \label{def-k}
\end{align}
where  $f$ is the \textit{gauge-invariant} two-form defined from the  gauge connection one-form $\mathbf{A}$   by 
\begin{align}
 f_{\alpha\beta}(x) 
  :=&  \partial_\alpha [\mathbf{n}(x) \cdot \mathbf{A}_\beta(x)] - 
  \partial_\beta [\mathbf{n}(x) \cdot \mathbf{A}_\alpha(x)] 
  \nonumber\\
  &+  ig^{-1} \mathbf{n}(x) \cdot [\partial_\alpha \mathbf{n}(x) \times  \partial_\beta \mathbf{n}(x)]  
   .
\end{align}
The  current $k$ is conserved in the sense that 
$\delta k=0$.
In $D=4$ dimensions, especially, we have
\begin{align}
 k_\mu = \frac12 \epsilon_{\mu\nu\alpha\beta} \partial_\nu f_{\alpha\beta}
 ,
\end{align}
and the magnetic charge $q_m$ is defined by 
\begin{align}
 q_m=\int d^3 \tilde{\sigma}_\mu k_\mu 
  ,
\end{align}
where $\bar{x}^\mu$ denotes a parameterization of the 3-dimensional volume $V$ and 
 $d^3 \tilde{\sigma}_{\mu} $ is the dual of the 3-dimensional volume element $d^3 \sigma^{\gamma_1\gamma_2\gamma_3}$:
\begin{align}
 d^3 \tilde{\sigma}_{\mu} 
:=  \frac{1}{3!}   \epsilon_{\mu\gamma_1\gamma_2\gamma_3} d^3 \sigma^{\gamma_1\gamma_2\gamma_3}
 , \quad
 d^3 \sigma^{\gamma_1\gamma_2\gamma_3}
 :=  \epsilon_{\beta_1\beta_2\beta_3} 
 \frac{\partial \bar{x}^{\gamma_1}}{\partial \sigma_{\beta_1}} \frac{\partial \bar{x}^{\gamma_1}}{\partial \sigma_{\beta_1}}  \frac{\partial \bar{x}^{\gamma_1}}{\partial \sigma_{\beta_1}}  d\sigma_{1} d\sigma_{2} d\sigma_{3} 
  .
\end{align}
See \cite{KSM08,Kondo08} for $SU(N)$ ($N \ge 3$) case.

The RDE in our reformulated Yang-Mills theory has the same form as that considered in BOT \cite{BOT97}, but its reasoning behind the RDE is quite different from the previous one, as can be seen from its derivation in  Appendix \ref{appendix:rc}, see \cite{KMS06} for more details.
We now give a new form of the RDE (eigenvalue-like equation): 
\begin{align}
 -D_\mu[\mathbf{A}]D_\mu[\mathbf{A}] \mathbf{n}(x) = \lambda(x) \mathbf{n}(x) 
 .
 \label{RDE2}
\end{align}
This implies that solving the RDE (\ref{RDE1}) is equivalent to look for the color field $\mathbf{n}(x)$ such that applying the covariant Laplacian $-D_\mu[\mathbf{A}]D_\mu[\mathbf{A}]$  of a given Yang-Mills field $\mathbf{A}_\mu(x)$  to the color field $\mathbf{n}(x)$ becomes parallel  to itself.  
It should be remarked that $\lambda(x)$ is non-negative, i.e., 
\begin{align}
 \lambda(x) \ge 0
 , 
\end{align}
since $-D_\mu[\mathbf{A}] D_\mu[\mathbf{A}]$ is a non-negative (positive definite) operator. 
The equivalence between (\ref{RDE1}) and (\ref{RDE2}) is shown as follows.
Let $\mathbf{e}_1(x), \mathbf{e}_2(x), \mathbf{e}_3(x) \equiv \bm{n}(x)$ be local orthonormal basis in SU(2) color space, i.e., 
\begin{align}
 \mathbf{e}_j(x) \times \mathbf{e}_k(x) = \epsilon_{jk\ell} \mathbf{e}_\ell(x) 
 , \quad
 \mathbf{e}_j(x) \cdot \mathbf{e}_k(x) = \delta_{jk} 
 .
\end{align}
Then we can write the left-hand side of (\ref{RDE2}) using three scalar functions $c_j(x)$ as
\begin{align}
 -D_\mu[\mathbf{A}]D_\mu[\mathbf{A}] \mathbf{n}(x) = c_1(x) \mathbf{e}_1(x) + c_2(x) \mathbf{e}_2(x) + c_3(x) \mathbf{e}_3(x)
 .
\end{align}
Now (\ref{RDE1}) is written as
\begin{align}
 \mathbf{0} =  \mathbf{n}(x) \times [-D_\mu[\mathbf{A}]D_\mu[\mathbf{A}] \mathbf{n}(x)] = c_1(x) \mathbf{e}_2(x) - c_2(x) \mathbf{e}_1(x)  
 .
\end{align}
By taking the inner product of both sides of this equation with $\mathbf{e}_1(x)$ or  $\mathbf{e}_2(x)$, we obtain $c_2(x) \equiv 0$ or $c_1(x) \equiv 0$, respectively.  Thus we obtain 
$
-D_\mu[\mathbf{A}]D_\mu[\mathbf{A}] \mathbf{n}(x) =  c_3(x) \mathbf{e}_3(x) =: \lambda(x) \bm{n}(x)
$.

An advantage of the new form (\ref{RDE2}) of RDE is as follows.
Once  the color field $\mathbf{n}(x)$ satisfying (\ref{RDE2}) is known, the value of the reduction functional $F_{\rm rc}$ is immediately calculable as an integral of the scalar function $\lambda(x)$ over the spacetime $\mathbb{R}^D$ as
\begin{align}
 F_{\rm rc} =&  \int d^Dx \frac12 (D_\mu[\mathbf{A}] \mathbf{n}(x)) \cdot (D_\mu[\mathbf{A}] \mathbf{n}(x)) 
 \nonumber\\
 =&   \int d^Dx \frac12  \mathbf{n}(x) \cdot (-D_\mu[\mathbf{A}] D_\mu[\mathbf{A}] \mathbf{n}(x)) 
  \nonumber\\
 =&  \int d^Dx \frac12  \mathbf{n}(x) \cdot \lambda(x) \mathbf{n}(x) 
  \nonumber\\
 =&  \int d^Dx  \frac12  \lambda(x)
 ,
 \label{Rc}
\end{align}
where we have used (\ref{unit}) in the last step.

Thus, the problem of solving the RDE has been reduced to another problem: 
 For a given Yang-Mills field $\mathbf{A}_\mu(x)$,  look for the unit vector field $\mathbf{n}(x)$ such that   $-D_\mu[\mathbf{A}] D_\mu[\mathbf{A}]\mathbf{n}(x)$ is proportional to $\mathbf{n}(x)$ 
with the smallest value of the reduction functional $F_{\rm rc}$ which is an integral of the scalar function $\lambda(x)$ over the spacetime $\mathbb{R}^D$.
For the integral (\ref{Rc}) to be convergent, $\lambda(x)$ must decrease rapidly for large $x^2$.


Our method should be compared with the conventional one of the  Laplacian Abelian gauge (LAG) \cite{BHVW01}.  LAG is performed by searching  for the field configuration $\phi$ which minimizes the functional
\begin{align}
 F_{\rm LAG} := \int d^Dx \frac12 (D_\mu[\mathbf{A}] \bm{\phi}(x))^2
 ,
 \label{LAGf}
\end{align}
provided that 
$\bm{\phi}(x)$ is square-integrable (and hence normalizable), i.e., 
\begin{align}
 \int d^Dx \bm{\phi}(x) \cdot \bm{\phi}(x) < \infty 
 .
 \label{sinteg}
\end{align}
The field configuration $\phi$ minimizing the functional can be viewed as the ground state with the lowest eigenvalue $\lambda$ in the eigenvalue equation for the covariant Laplacian $-D_\mu[\mathbf{A}]D_\mu[\mathbf{A}]$:
\begin{align}
 -D_\mu[\mathbf{A}]D_\mu[\mathbf{A}] \bm{\phi}(x) = \lambda  \bm{\phi}(x) 
 .
 \label{LAG}
\end{align}
Once the eigenvalue problem is solved, the functional reads
\begin{align}
 F_{\rm LAG} 
:=& \int d^Dx  \frac12 \bm{\phi}(x) \cdot [-D_\mu[\mathbf{A}] D_\mu[\mathbf{A}] \bm{\phi}(x) ] 
 \nonumber\\
 =& \frac12 \lambda   \int d^Dx  \bm{\phi}(x) \cdot \bm{\phi}(x)   
= \frac12 \lambda      
 .
 \label{LAGf1}
\end{align}
Therefore, the lowest eigenvalue $\lambda$ gives the smallest value of the LAG functional $F_{\rm LAG}$. 
It should be remarked that $\lambda$ is non-negative, i.e., $\lambda \ge 0$, since $-D_\mu[\mathbf{A}] D_\mu[\mathbf{A}]$ is a non-negative (positive definite) operator. 
In LAG, the variable $\lambda$ is regarded as the Lagrange multiplier for incorporating a constraint (\ref{sinteg}) in the minimization problem: 
\begin{align}
 F_{\rm LAG} := \int d^Dx \frac12 (D_\mu[\mathbf{A}] \bm{\phi}(x))^2 - \lambda \left[ \int d^Dx \frac12 \bm{\phi}(x) \cdot \bm{\phi}(x)  -1 \right]
 .
 \label{LAGf2}
\end{align}
This is not the case for the reduction functional $F_{\rm rc}$.

\section{Simplifying the RDE using symmetries}

In what follows, we restrict our considerations to  the four-dimensional $D=4$ Euclidean Yang-Mills theory. 

\subsection{CFtHW Ansatz for Yang-Mills field}

First, we adopt the Corrigan-Fairlie-'tHooft-Wilczek (CFtHW) Ansatz \cite{CF77,tHooft76,Wilczek77}:
\begin{align}
 g\mathbf{A}_\mu(x) 
=   \frac{\sigma_A}{2} g\mathbf{A}_\mu^A(x) 
=  \frac{\sigma_A}{2} \eta^A_{\mu\nu} f_\nu(x) , \quad f_\nu(x) := \partial_\nu \ln \Phi(x) 
 ,
\end{align}
where $\eta^A_{\mu\nu}=\eta^{(+)}{}^A_{\mu\nu}$ is the symbol defined by 
\begin{align}
 \eta^A_{\mu\nu} \equiv \eta^{(+)}{}^A_{\mu\nu} := \epsilon_{A\mu\nu 4} + \delta_{A\mu}\delta_{\nu 4} - \delta_{\mu 4}\delta_{A \nu}
 = \begin{cases}
    \epsilon_{Ajk} & (\mu=j, \nu=k) \cr
    \delta_{Aj} & (\mu=j, \nu= 4) \cr
    -\delta_{Ak} & (\mu=4, \nu=k)
   \end{cases}
 .
\end{align}
Similarly, we can define
$\bar{\eta}^A_{\mu\nu}=:\eta^{(-)}{}^A_{\mu\nu}$  as
\begin{align}
 \bar{\eta}^A_{\mu\nu} \equiv  \eta^{(-)}{}^A_{\mu\nu} := \epsilon_{A\mu\nu 4} - \delta_{A\mu}\delta_{\nu 4} + \delta_{\mu 4}\delta_{A \nu}
 = \begin{cases}
    \epsilon_{Ajk} & (\mu=j, \nu=k) \cr
    -\delta_{Aj} & (\mu=j, \nu= 4) \cr
    +\delta_{Ak} & (\mu=4, \nu=k)
   \end{cases}
 .
\end{align}
Note that $\eta^A_{\mu\nu}$ is self-dual, i.e., $\eta^A_{\mu\nu}=*\eta^A_{\mu\nu}:=\frac12 \epsilon_{\mu\nu\alpha\beta} \eta^A_{\alpha\beta}$,
while $\bar{\eta}^A_{\mu\nu}$ is anti-selfdual, i.e., 
$-\bar{\eta}^A_{\mu\nu}=*\bar{\eta}^A_{\mu\nu}$.

Under this Ansatz, the RDE is greatly simplified:
\begin{align}
 \{ [-\partial_\mu \partial_\mu + 2 f_\mu f_\mu ]\delta_{AB} + 2 \epsilon_{ABC} \eta^C_{\mu\nu} f_\nu(x) \partial_\mu \} \mathbf{n}_B(x) = \lambda(x) \mathbf{n}_A(x)  
 .  
 \label{RDE0}
\end{align}
See Appendix \ref{appendix:RDE} for the derivation.

The Yang-Mills field in the  CFtHW  Ansatz satisfies simultaneously the Lorentz gauge:
\begin{align}
 \partial_\mu \mathbf{A}_\mu^A(x)= 0
 ,
\end{align}
and the maximal Abelian gauge (MAG): 
\begin{align}
D_\mu[\mathbf{A}^3] \mathbf{A}_\mu^{\pm}(x):= (\partial_\mu - ig \mathbf{A}_\mu^3) (\mathbf{A}_\mu^1(x) \pm i \mathbf{A}_\mu^2(x)) = 0
 .
\end{align}

\subsection{Symmetry}

In order to further simplify the equation,  we make use of the Euclidean rotation group $SO(4)$.
This symmetry enables one to separate the RDE  into the angular and radial parts. 
We define the generators of four-dimensional Euclidean rotations as 
\begin{align}
  L_{\mu\nu} = -i(x_\mu \partial_\nu - x_\nu \partial_\mu) 
   , \quad \mu, \nu \in \{ 1, 2, 3, 4 \}
  .
   \label{Lmn}
\end{align}
Indeed, it is straightforward to check that $L_{\mu\nu}$ satisfies the Lie algebra of $SO(4)$.  The angular part is expressed in terms of angular momentum derived from the decomposition:
\begin{align}
  so(4) \cong su(2) + su(2)
   .
\end{align}
In analogy with the Lorentz group, one introduces the angular momentum and boost generators:
\begin{align}
  \mathscr{L}_j := \frac12 \epsilon_{jk\ell} L_{k\ell} , 
  \quad
  \mathscr{K}_j := L_{j4},  \quad  j,k,\ell \in \{ 1,2,3 \} 
   ,
   \label{LjKj}
\end{align}
and their linear combinations:
\begin{align}
 M_A :=& \frac12 (\mathscr{L}_A-\mathscr{K}_A) = - \frac{i}{2} \bar{\eta}^A_{\mu\nu} x_\mu \partial_\nu ,
 \quad  A \in \{ 1,2,3 \} 
 ,
 \nonumber\\
 N_A :=& \frac12 (\mathscr{L}_A+\mathscr{K}_A) = - \frac{i}{2}  \eta^A_{\mu\nu} x_\mu \partial_\nu , 
\quad A \in \{ 1,2,3 \} 
 .
\end{align}
The operators $M_A$ and $N_A$ generate two independent $SU(2)$ subgroups with Casimir operators $\vec{M}^2 :=M_A M_A$ and $\vec{N}^2 :=N_A N_A$ having eigenvalues $M(M+1)$ and $N(N+1)$, respectively:
\begin{align}
 \vec{M}^2 :=& M_A M_A \rightarrow M(M+1), \quad  M \in \{ 0, \frac12, 1, \frac32, \cdots \} 
  ,
 \nonumber\\
 \vec{N}^2 :=& N_A N_A \rightarrow N(N+1), \quad  N \in \{ 0, \frac12, 1, \frac32, \cdots \} 
 .
\end{align}
Here it is important to note that the eigenvalues $M$ and $N$ are half-integers. 

The generators for isospin $S=1$ are 
\begin{align}
 (S_A)_{BC} := i\epsilon_{ABC} = (S_C)_{AB}  
 .
\end{align}
It is easy to see that $\vec{S}^2$ is a Casimir operator
and
$\vec{S}^2$ has the eigenvalue
\begin{align}
 \vec{S}^2 := S_A S_A \rightarrow S(S+1) = 2
 , 
\end{align}
since   
\begin{align}
 (\vec{S}^2)_{AB} = (S_C)_{AD} (S_C)_{DB} 
= i\epsilon_{DCA} i\epsilon_{BCD} = 2 \delta_{AB} 
 .
\end{align}

Now we introduce the conserved total angular momentum $\vec{J}$ by
\begin{align}
 \vec{J} = \vec{L} + \vec{S}
 ,  
\end{align}
with the eigenvalue 
\begin{align}
 \vec{J}^2 \rightarrow J(J+1), \quad J \in \{ L+1, L, |L-1| \} 
 ,
\end{align}
where $\vec{L}=\vec{M}$ or $\vec{L}=\vec{N}$.
Using the representations (\ref{Lmn}) and (\ref{LjKj}),  we find  that 
\begin{align}
 \vec{N}^2 -\vec{M}^2 = 0 = \vec{\mathscr{L}} \cdot \vec{\mathscr{K}} 
 .
\end{align}
Thus, a complete set of commuting observables is given by the Casimir operators, $\vec{J}^2$, $\vec{L}^2$, $\vec{S}^2$ and their projections, e.g., $J_z, L_z, S_z$.  

By using
\begin{align}
 \vec{S} \cdot \vec{L} = (\vec{J}^2-\vec{L}^2-\vec{S}^2)/2 
 ,
\end{align}
the RDE is rewritten in the form:
\begin{align}
 \{ -\partial_\mu \partial_\mu \delta_{AB} + 2f(x) (\vec{J}^2-\vec{L}^2-\vec{S}^2)_{AB} + x_\mu x_\mu f^2(x) (\vec{S}^2)_{AB} \} \mathbf{n}_B(x) = \lambda(x) \mathbf{n}_A(x) 
 ,
\end{align}
where the spherical symmetry  allows us to take 
\begin{align}
  f_\nu(x) := \partial_\nu \ln \tilde{\Phi}(x^2) = x_\nu f(x) 
 .
\end{align}

The symmetry consideration suggests that $\mathbf{n}(x)$ is separated into the radial and angular part: In the vector (component) form:
\begin{subequations}
\begin{align}
 \mathbf{n}_A(x) = \psi(R) Y^A_{(J,L)}(\hat{x}) 
 ,
\end{align}
or in the Lie algebra valued form:
\begin{align} 
 \bm{n}(x) =& \mathbf{n}_A(x) \sigma_A =  \psi(R) Y^A_{(J,L)}(\hat{x}) \sigma_A 
 ,
\nonumber\\
 & R := \sqrt{x_\mu x_\mu} \in \mathbb{R}_{+}, 
 \quad \hat{x}_\mu:=x_\mu/R \in S^3
\end{align}
\end{subequations}
where
$\vec{Y}_{(J,L)}(\hat{x})=\{ Y^A_{(J,L)}(\hat{x}) \}_{A=1,2,3}$ denote the vector spherical harmonics on $S^3$ characterized by 
\begin{align}
 \vec{L}^2 Y^A_{(J,L)}(\hat{x}) =& L(L+1) Y^A_{(J,L)}(\hat{x})
,
 \\
 \vec{J}^2 Y^A_{(J,L)}(\hat{x}) =& J(J+1) Y^A_{(J,L)}(\hat{x})
,
 \\
 \vec{S}^2 Y^A_{(J,L)}(\hat{x}) \sigma_A =& S(S+1) Y^A_{(J,L)}(\hat{x}) \sigma_A 
 ,
\end{align}
with $S=1$.  
The explicit form of the vector spherical harmonics is given later.

In this form, the covariant Laplacian reduces to the diagonal form and RDE reduces to 
\begin{align}
 & [ -\partial_\mu \partial_\mu  + V(x)  ] \mathbf{n}_A(x) = \lambda(x) \mathbf{n}_A(x) 
 ,
\nonumber\\
 & V(x) :=  2f(x) [J(J+1)-L(L+1)-2] + 2x^2 f^2(x)  
 .
 \label{eq1}
\end{align}
This equation does not necessarily mean that the left-hand side of the RDE becomes automatically proportional to $\mathbf{n}(x)$, since $\partial_\mu \partial_\mu \mathbf{n}(x)$ is not guaranteed to be proportional to $\mathbf{n}(x)$. If this is the case, we have
\begin{align}
 \lambda(x) = V(x) + [-\partial_\mu \partial_\mu \mathbf{n}_A(x)]/\mathbf{n}_A(x) 
 \quad \text{for any A, no sum over A}
 .
\end{align}

Moreover, it is possible to rewrite the Laplacian in terms of the radial coordinate $R$ and the angular coordinates:
\begin{align}
 -\partial_\mu \partial_\mu  
 = - \partial_R \partial_R  - \frac{3}{R} \partial_R + \frac{2(\vec{M}^2+\vec{N}^2)}{R^2} 
 , \quad R := \sqrt{x_\mu x_\mu} 
 ,
\end{align}
which reads
\begin{align}
 -\partial_\mu \partial_\mu  
 = - \partial_R \partial_R  - \frac{3}{R} \partial_R + \frac{4 \vec{L}^2}{R^2} 
  = - \frac{1}{R^3} \frac{\partial}{\partial R} \left( R^3 \frac{\partial}{\partial R}  \right) + \frac{4 \vec{L}^2}{R^2} 
 .
\end{align}
Thus, we arrive at another expression of RDE:
\begin{align}
  \left[ - \partial_R \partial_R  - \frac{3}{R} \partial_R    + \tilde{V}(x)  \right] \mathbf{n}_A(x)  
&=  \lambda(x) \mathbf{n}_A(x) 
 ,
\nonumber\\
  \tilde{V}(x) 
:=&   \frac{4L(L+1)}{x^2}  + V(x) 
\nonumber\\
 =&   \frac{4L(L+1)}{x^2}  + 2f(x) [J(J+1)-L(L+1)-2] + 2x^2 f^2(x)  
 .
\end{align}
If the left-hand side of the RDE becomes proportional to $\mathbf{n}(x)$, $\lambda(x)$ is given by
\footnote{
The second term of the right-hand side of (\ref{eq2}) does not contribute to $\lambda(x)$ if and only if 
$
\psi(R)=C_1+C_2/R^2
$.
If $\psi(R) \sim R^{-\gamma}$, then the second term contributes $\gamma(2-\gamma)/R^2$.  This increases $\lambda(x)$ for $0<\gamma<2$, while it decreases $\lambda(x)$ for $\gamma<0$ and $\gamma >2$.
}
\begin{align}
 \lambda(x) = \tilde{V}(x)  - \psi(R)^{-1} \frac{1}{R^3} \frac{\partial}{\partial R} \left( R^3 \frac{\partial}{\partial R} \psi(R) \right)
 .
 \label{eq2}
\end{align}

\subsection{Unit vector condition and angular part}

In rewriting RDE due to $SO(4)$ symmetry,  
 we have not yet used the fact that $\mathbf{n}(x)$ has the unit length: 
\begin{align}
 1 = \mathbf{n}(x) \cdot \mathbf{n}(x) = \mathbf{n}_A(x)  \mathbf{n}_A(x) 
= \psi(R) \psi(R)  Y^A_{(J,L)}(\hat{x})  Y^A_{(J,L)}(\hat{x}) 
 .
\end{align}
If the vector spherical harmonics happens to be normalized at every spacetime point as
\begin{align}
 1 =  Y^A_{(J,L)}(\hat{x})  Y^A_{(J,L)}(\hat{x}) 
 ,
 \label{YY}
\end{align}
then we can take without loss of generality
\begin{align}
 \psi(R) \equiv 1 
 ,
\end{align}
and $\mathbf{n}(x)$ is determined  by the vector spherical harmonics alone: 
\begin{align}
 \mathbf{n}_A(x)   
=  Y^A_{(J,L)}(\hat{x})   
 .
\end{align}
The degeneracy of the state $Y^A_{(J,L)}$ is given by $(2J+1)(2L+1)$.  
In this case, the lowest value of $\lambda(x)$ is obtained by minimizing $\tilde{V}(x)$ at every $x$. 
However, (\ref{YY}) is not guaranteed for any set of $(J,L)$ except for some special cases, as we see shortly. 
\footnote{
Usually, the orthonormality of the vector spherical harmonics is given with respect to the integral over $S^3$ with a finite volume:
\begin{align}
 \int_{S^3} d\Omega \ Y^A_{(J,L)}(\hat{x})  Y^A_{(J',L')}(\hat{x}) = \delta_{JJ'} \delta_{LL'} 
 .
\end{align}
}

\section{One-instanton case}

In order to treat meron and instanton (in the regular gauge) simultaneously, we adopt the form:
\begin{align}
 f_\mu(x) = x_\mu f(x), \quad f(x) = \frac{2\kappa}{x^2+s^2}
 . 
\end{align}
For a given set of $(J,L)$, we have calculated the ``potential'' $\tilde{V}(x)$ and the ``eigenvalue'' $\lambda(x)$, which  are enumerated in the following 
Table. 
 Note that $(J,L)=(0,0)$ is excluded by selection rules for $S=1$.

\begin{center}
\begin{tabular}{l||l|l||l|l|l} \hline\hline
J & L & S & degeneracy & 1-instanton (zero size) & 1-meron \\  
  &   &   &   & $\tilde{V}(x)$ & $\tilde{V}(x)$ \\ \hline\hline
1 & 0 & 1     & 3  & $8/x^2$ & $2/x^2$ \\ \hline
1/2 & 1/2 & 1 & 4  & $3/x^2$ & $1/x^2$ \\  
3/2 & 1/2 & 1 & 8  & $15/x^2$ & $7/x^2$ \\ \hline
0 & 1 & 1     & 3  & $0$ & $2/x^2$ \\ 
1 & 1 & 1     & 9  & $8/x^2$ & $6/x^2$ \\
2 & 1 & 1     & 15 & $24/x^2$ & $14/x^2$ \\ \hline
\end{tabular}
\label{table:eigenvalue1}
\end{center}

\subsection{One instanton in the regular (or non-singular) gauge}

One-instanton configuration in the regular gauge with zero size, i.e., $\kappa=1$, $s=0$, is expressed by
\begin{align}
 f(x) = \frac{2}{x^2} 
 ,
\end{align}
which leads to 
\begin{align}
 V(x) =  \frac{4}{x^2} [J(J+1)-L(L+1)]   
 , \quad
 \tilde{V}(x) =  \frac{4}{x^2} J(J+1)  \ge 0
 .
\end{align}
For one-instanton with zero size in the regular gauge, therefore, $(J,L)=(0,1)$ gives the lowest value of $\tilde{V}(x)$ at every $x$.
Hence the lowest value of $\lambda(x)$ is obtained  $\lambda(x)=\tilde{V}(x)=0$ if we can set $\psi(R) \equiv {\rm const.}$ from (\ref{eq2}).  This is the lowest possible  value, since $\lambda(x) \ge 0$. For this to be satisfied, the corresponding vector harmonics must be orthonormal (\ref{YY}). 
The vector spherical harmonics $Y_{(0,1)}(\hat{x})$ is 3-fold degenerate and is written as a linear combination of three degenerate states  ($B=1,2,3)$:
 \footnote{
This degeneracy corresponds to the Gribov copies associated with the reduction (partial) gauge fixing from the enlarged gauge symmetry $SU(2)\times SU(2)/U(1)$ to the original gauge symmetry $SU(2)$, see \cite{KMS06}.  
These Gribov copies are true Gribov copies, but are different from those in fixing the original gauge symmetry $SU(2)$.
}
\begin{align}
& \bm{Y}_{(0,1)}(\hat{x})
 \nonumber\\
=& \sum_{B=1}^{3} \hat{a}_B \bm{Y}_{(0,1),(B)}(\hat{x})
 \nonumber\\
=& \hat{a}_1 
 \begin{pmatrix}
 \hat{x}_1^2-\hat{x}_2^2-\hat{x}_3^2+\hat{x}_4^2  \cr
 2(\hat{x}_1\hat{x}_2-\hat{x}_3\hat{x}_4) \cr
 2(\hat{x}_1\hat{x}_3+\hat{x}_2\hat{x}_4)
 \end{pmatrix}
 + \hat{a}_2 
 \begin{pmatrix}
 2(\hat{x}_1\hat{x}_2+\hat{x}_3\hat{x}_4)  \cr
  -\hat{x}_1^2+\hat{x}_2^2-\hat{x}_3^2+\hat{x}_4^2 \cr
 2(\hat{x}_2\hat{x}_3-\hat{x}_1\hat{x}_4)
 \end{pmatrix}
 + \hat{a}_3 
 \begin{pmatrix}
 2(\hat{x}_1\hat{x}_3-\hat{x}_2\hat{x}_4) \cr
 2(\hat{x}_2\hat{x}_3+\hat{x}_1\hat{x}_4) \cr
 -\hat{x}_1^2-\hat{x}_2^2+\hat{x}_3^2+\hat{x}_4^2
 \end{pmatrix}
 ,
\end{align}
where  $\hat{a}_B$ are coefficients of the linear combination.  Hereafter the vector with the hat symbol denotes a unit vector, e.g., $\hat{a}_B\hat{a}_B=1$.
It is easy to check that $Y_{(0,1)}(\hat{x})$ are orthonormal at every point:
\begin{align}
\bm{Y}_{(0,1),(B)}(\hat{x}) \cdot \bm{Y}_{(1,0),(C)}(\hat{x}) 
:=  Y^A_{(0,1),(B)}(\hat{x}) Y^A_{(1,0),(C)}(\hat{x})
  = \delta_{BC} 
 .
\end{align}
Thus the  solution is given by the linear combination of triplet of vector spherical harmonics $\bm{Y}_{(0,1)}(\hat{x})$, which is written in the manifestly Lorentz covariant Lie-algebra valued form using 
  Pauli matrices $\sigma_A$ and 
\begin{align}
 \bar{e}_\mu = (i\sigma_A, \mathbf{1}), 
 \quad
 e_\mu := (-i\sigma_A, \mathbf{1}) 
 ,
\end{align}
as 
\begin{align}
 \bm{n}(x) := \mathbf{n}_A(x) \sigma_A 
= \hat{a}_B  Y^A_{(0,1),(B)}(\hat{x}) \sigma_A 
= \hat{a}_B x_\alpha \bar{e}_\alpha \sigma_B x_\beta e_\beta/x^2 
 ,
 \label{Hopf-sol}
\end{align}
or in the vector component 
\begin{align}
  \mathbf{n}_A(x) 
= \hat{a}_B  Y^A_{(0,1),(B)}(\hat{x}) 
=  \hat{a}_B x_\alpha x_\beta \bar{\eta}^B_{\alpha \gamma} \eta^A_{\gamma\beta}/x^2
 .
 \label{hedgehog-sol2}
\end{align}
where we have used the formula:
\begin{align}
 {\rm tr}[\sigma_A \bar{e}_\alpha \sigma_B e_\beta  ]
 = -2 \bar{\eta}^B_{\alpha \gamma} \eta^A_{\beta\gamma} 
 .
\end{align}

It is directly checked that (\ref{Hopf-sol}) is indeed the solution of the RDE. 
Explicit calculations show  that (\ref{Hopf-sol}) satisfies 
\begin{align}
  -\partial_\mu \partial_\mu \mathbf{n}_A(x) = \frac{8}{x^2} \mathbf{n}_A(x)
 ,
\end{align}
and
\begin{align}
  2 \epsilon_{ABC} \eta^C_{\mu\nu} f_\nu(x) \partial_\mu  \mathbf{n}_B(x)
  = - 8 f(x) \mathbf{n}_A(x) 
  = - \frac{16}{x^2} \mathbf{n}_A(x) 
 .
\end{align}
Then, for $(J,L)=(0,1)$, we arrive at  
\begin{align}
 V(x) =  \frac{-8}{x^2} 
 , \quad
 \tilde{V}(x) = 0 
 ,
\end{align}
and
\begin{align}
 \lambda(x) = V(x) + [-\partial_\mu \partial_\mu \mathbf{n}_A(x)]/\mathbf{n}_A(x)
\equiv 0 
 \quad \text{for any A, no sum over A}
 .
\end{align}
Thus this solution is an allowed one, since the solution gives a finite (vanishing) value for the functional $F_{\rm rc}$=0.  
The solution gives a map $\bm{Y}_{(0,1),(B)}$ from $S^3$ to $S^2$, which is known as the standard Hopf map.  
Therefore, the only zeros of $\phi_A(x)$ in the solution
$\mathbf{n}_A(x) =\phi_A(x)/|\phi(x)|=\phi_A(x)/\sqrt{\phi_B(x)\phi_B(x)}$
are the origin and the set of magnetic monopoles consists of the origin only, in other words, the magnetic monopole loop is shrank to a single point. 
Therefore, we have no  monopole loop with a finite and non-zero radius for the Yang-Mills field of one instanton with zero size in the regular gauge.

For one instanton with  size $\rho$, i.e., $\kappa=1$, $s=\rho$, we must examine
\begin{align}
 f(x^2) = \frac{2}{x^2+\rho^2}
 ,
\end{align}
and
\begin{align}
 V(x) =  \frac{4}{x^2+\rho^2} [J(J+1)-L(L+1) ]  - \frac{8\rho^2}{(x^2+\rho^2)^2} 
 .
\end{align}
The lowest $\lambda(x)$ is realized for distinct set of $(J,L)$  depending on the region of $x$. 
This case is obtained by one-instanton limit of two meron case to be discussed later.

\subsection{One instanton in the singular gauge}

For one instanton in the singular gauge, we must take
\begin{align}
g\mathbf{A}_\mu(x) 
=  \frac{\sigma_A}{2} \bar{\eta}^A_{\mu\nu} x_\nu f(x^2) ,\quad
 f(x^2) = \frac{2\rho^2}{x^2(x^2+\rho^2)}
 .
\end{align}
The results in the previous section hold by replacing $\eta^A_{\mu\nu}$ by $\bar{\eta}^A_{\mu\nu}$.  
In this case, we have 
\begin{align}
 V(x) =  \frac{4\rho^2}{x^2(x^2+\rho^2)} [J(J+1)-L(L+1)-2] +   \frac{8\rho^4}{x^2(x^2+\rho^2)^2}  
 .
\end{align}
Apart from the detailed analysis, we focus on the zero size limit $\rho \rightarrow 0$ (or the distant region $x^2 \rightarrow \infty$): 
\begin{align}
 V(x) \simeq 0
 , \quad
 \tilde{V}(x) \simeq \frac{4L(L+1)}{x^2}   
 .
\end{align}
It is easy to see that the solution is given at $(J,L)=(1,0)$, i.e.,  
$\mathbf{n}(x)=\bm{Y}_{(1,0)}$ (a constant vector) given in (\ref{Y10}), which has 
the lowest value of $\lambda(x)$:  $\lambda(x) \equiv 0$. 
For  $(J,L)=(1,0)$, the state is 3-fold degenerate:  $\mathbf{n}(x)=\bm{Y}_{(1,0)}$ is written as a linear combination of them:
Writing $\bm{Y}_{(1,0)}$ as a column vector: 
$\bm{Y}_{(1,0)}=(Y^1_{(1,0)},Y^2_{(1,0)},Y^3_{(1,0)})^T$ (T denotes transpose)
\begin{align}
 \bm{Y}_{(1,0)} 
 = \sum_{\alpha=1}^{3} \hat{c}_\alpha \bm{Y}_{(1,0),(\alpha)}
= \hat{c}_1 
 \begin{pmatrix}
 1 \cr
 0 \cr
 0 
 \end{pmatrix}
 + \hat{c}_2 
 \begin{pmatrix}
 0 \cr
 1 \cr
 0 
 \end{pmatrix}
 + \hat{c}_3 
 \begin{pmatrix}
 0 \cr
 0 \cr
 1 
 \end{pmatrix}
 .
 \label{Y10}
\end{align}
It constitutes the orthonormal set:
\begin{align}
\bm{Y}_{(1,0),(\alpha)} \cdot \bm{Y}_{(1,0),(\beta)} 
:=  Y^A_{(1,0),(\alpha)} Y^A_{(1,0),(\beta)}  = \delta_{\alpha \beta} 
 . 
\end{align}
Therefore, 
the solution is given by a constant:
\begin{align}
 \mathbf{n}_A(x) 
= \sum_{\alpha=1}^{3} \hat{c}_\alpha Y^A_{(1,0),(\alpha)} 
= \hat{c}_A  
 .
\end{align}
In this limit, 
$
  \partial_\mu \mathbf{n}_A(x) =0
$,
$
 \partial_\mu \partial_\mu \mathbf{n}_A(x) =0
$ 
and 
\begin{align}
 \lambda(x) = V(x)  
=   2x^2 f^2(x)  
= \frac{8\rho^4}{x^2(x^2+\rho^2)^2}
 .
\end{align}
One-instanton in the singular gauge yields a finite reduction functional: 
\begin{align}
 F_{\rm rc} = \int d^4x  \lambda(x)  < \infty
 .
\end{align}

\section{One-meron and magnetic monopole line}

In order discuss one-meron configuration, i.e.,  $\kappa=\frac12$, $s=0$, we have
\begin{align}
 f(x^2) = \frac{1}{x^2} 
 ,
\end{align}
which yields 
\begin{align}
 V(x) =  \frac{2}{x^2}  [J(J+1)-L(L+1)-1]   
 , \quad
 \tilde{V}(x) =  \frac{2}{x^2} [J(J+1)+L(L+1)-1]   > 0
 .
\end{align}
For one meron, we find that $(J,L)=(1/2,1/2)$ gives the lowest $\tilde{V}(x)$. 
This suggests that the solution might be given by 
\begin{align}
 \bm{Y}_{(1/2,1/2)}(\hat{x}) 
 &= \sum_{\mu=1}^{4} \hat{b}_\mu \bm{Y}_{(1/2,1/2),(\mu)}(\hat{x})
\nonumber\\
&= \hat{b}_1 
 \begin{pmatrix}
 -\hat{x}_4 \cr
 \hat{x}_3 \cr
 -\hat{x}_2 
 \end{pmatrix}
 + \hat{b}_2 
 \begin{pmatrix}
 -\hat{x}_3 \cr
 -\hat{x}_4 \cr
 \hat{x}_1 
 \end{pmatrix}
 + \hat{b}_3 
 \begin{pmatrix}
 \hat{x}_2 \cr
 -\hat{x}_1 \cr
 -\hat{x}_4 
 \end{pmatrix}
 + \hat{b}_4 
 \begin{pmatrix}
 \hat{x}_1 \cr
 \hat{x}_2 \cr
 \hat{x}_3 
 \end{pmatrix}
 ,
 \label{Y-hedgehog}
\end{align}
where a unit four-vector $\hat{b}_\mu$ ($\mu=1,2,3,4$) denote four coefficients of the linear combination for 4-fold generate $\bm{Y}_{(1/2,1/2),(\mu)}(\hat{x})$ ($\mu=1,2,3,4$).

However, a subtle point in this case is that $Y^A_{(1/2,1/2),(\mu)}(\hat{x})$ are non-orthonormal sets at every spacetime point: 
\begin{align}
\bm{Y}_{(1/2,1/2),(\mu)}(\hat{x}) \cdot \bm{Y}_{(1/2,1/2),(\nu)}(\hat{x}) 
:=  Y^A_{(1/2,1/2),(\mu)}(\hat{x}) Y^A_{(1/2,1/2),(\nu)}(\hat{x})
  \ne \delta_{\mu\nu} 
 .
\end{align}

Nevertheless, we find that the unit vector field:
\begin{align}
 \mathbf{n}_A(x) 
= b_\nu \eta^A_{\mu\nu} x_\mu /\sqrt{b^2 x^2 -(b \cdot x)^2}
= \hat{b}_\nu \eta^A_{\mu\nu} \hat{x}_\mu /\sqrt{ 1 -(\hat{b} \cdot \hat{x})^2}
 ,
 \label{n-hedgehog-sol}
\end{align}
constructed from 
\begin{align}
 \bm{Y}_{(1/2,1/2),(\mu)}(\hat{x})=\eta^A_{\mu\nu} \hat{x}_\nu  \quad ( \mu=1,2,3,4 )
 ,
\end{align}
can be a solution of RDE. 
In fact, explicit calculations show  that (\ref{n-hedgehog-sol}) satisfies
\begin{align}
  -\partial_\mu \partial_\mu \mathbf{n}_A(x) = \frac{2}{ x^2-(\hat{b} \cdot x)^2} \mathbf{n}_A(x)
 ,
\end{align}
and
\begin{align}
  2 \epsilon_{ABC} \eta^C_{\mu\nu} f_\nu(x) \partial_\mu  \mathbf{n}_B(x)
  = - 4 f(x) \mathbf{n}_A(x) 
  = - \frac{4}{x^2} \mathbf{n}_A(x) 
 .
\end{align}
Then, for $(J,L)=(1/2,1/2)$, we conclude that 
\begin{align}
 V(x) =  \frac{-2}{x^2}  
 , \quad
 \tilde{V}(x) = \frac{1}{x^2} 
 ,
\end{align}
and
\begin{align}
 \lambda(x) =& [-\partial_\mu \partial_\mu \mathbf{n}_A(x)]/\mathbf{n}_A(x) + V(x) 
 \quad \text{for any A, no sum over A}
 \nonumber\\
 =&  \frac{2(\hat{b} \cdot x)^2}{x^2[ x^2-(\hat{b} \cdot x)^2]}
 .
 \label{meron-eigen}
\end{align}
The solution (\ref{n-hedgehog-sol}) is of the hedgehog type.
The magnetic monopole current is obtained as simultaneous zeros of $\hat{b}_\nu \eta^A_{\mu\nu}x_\mu=0$ for $A=1,2,3$. 
Taking the 4th vector in (\ref{Y-hedgehog}) $\hat{b}_\mu= \delta_{\mu 4}$, the magnetic monopole current is located at $x_1=x_2=x_3=0$, i.e., on the $x_4$ axis. Whereas, if the 3rd vector in (\ref{Y-hedgehog}) is taken $\hat{b}_\mu= \delta_{\mu 3}$, the magnetic  monopole current flows at $x_1=x_2=x_4=0$, i.e., on $x_3$ axis.
In general, it turns out that the magnetic monopole current $k_\mu$  is located on the straight line parallel  to $\hat{b}_\mu$ going through the origin. 
Note that the expression (\ref{meron-eigen}) for $\lambda(x)$ is invariant under a subgroup $SO(3)$ of  the Euclidean rotation $SO(4)$. 
In other words, once we select $\hat{b}_\mu$, $SO(4)$ symmetry is broken to $SO(3)$ just as in the spontaneously broken symmetry.  This result is consistent with a fact that the magnetic monopole current $k_\mu$ flows in the direction of $\hat{b}_\mu$ and the symmetry is reduced to the axial symmetry, the rotation group $SO(3)$, about the axis  in the direction of a four vector $\hat{b}_\mu$.

It is instructive to point out that the Hopf map $Y$ also satisfies the RDE.  
Therefore, it is necessary to compare the value of the reduction functional of $(J,L)=(1/2,1/2)$ with that of $(J,L)=(0,1)$. 
In the $(J,L)=(0,1)$ case, we find
\begin{align}
  \lambda_{(0,1)}(x)
 =   \frac{2}{x^2} = \frac{2}{x_1^2+x_2^2+x_3^2+x_4^2}
 .
\end{align}
For instance, we can choose $\hat{b}_\mu= \delta_{\mu 3}$ without loss of generality: 
\begin{align}
  \lambda_{(1/2,1/2)}(x)
 =   \frac{2x_3^2}{[x_1^2+x_2^2+x_3^2+x_4^2][ x_1^2+x_2^2+x_4^2]}
 .
\end{align}
Note that  the integral of $\lambda_{(1/2,1/2)}(x)$ over the whole spacetime $\mathbb{R}^4$ is obviously smaller than that of $\lambda_{(0,1)}(x)$, although 
$\lambda_{(0,1)}(x) < \lambda_{(1/2,1/2)}(x)$ locally 
inside a cone with the symmetric axis $\hat{b}_\mu$, i.e.,
$(\hat{b} \cdot \hat{x})^2 \ge 1/2$.

The reduction functional in $(J,L)=(1/2,1/2)$ case reads 
\begin{align}
 F_{\rm rc} 
 =&  \int d^4x  \frac12 \lambda_{(1/2,1/2)}(x)
 = \int  dx_3 \int dx_1dx_2dx_4  \frac{x_3^2}{[x_1^2+x_2^2+x_3^2+x_4^2][ x_1^2+x_2^2+x_4^2]}
 \nonumber\\
 =& 4\pi \int_{-L_3}^{L_3} dx_3 x_3^2 \int_{0}^{\infty} dr  \frac{1}{[r^2+x_3^2]}
 \nonumber\\
 =& 4\pi \int_{-L_3}^{L_3} dx_3 x_3^2  \frac{1}{|x_3|} \arctan \frac{r}{|x_3|} \Big|_{r=0}^{r=\infty}
 \nonumber\\
 =& 4\pi \int_{-L_3}^{L_3} dx_3 x_3^2  \frac{1}{|x_3|}  \frac{\pi}{2} 
 \nonumber\\
 =& 4\pi^2 \int_{0}^{L_3} dx_3  x_3  
 ,
\end{align}
where we have defined $r^2:=x_1^2+x_2^2+x_4^2$.
$\lambda_{(1/2,1/2)}(x)$ is zero on the $x_3=0$ hyperplane, i.e.,  three dimensional space $x_1,x_2,x_4$ which is orthogonal to the magnetic current. Therefore, in the three-dimensional space, the magnetic current looks like just a point magnetic charge.

Although $F_{\rm rc}$ remains finite as long as $L_3$ is finite,  it diverges for $L_3 \rightarrow \infty$, i.e, when integrated out literally in the whole spacetime $\mathbb{R}^4$.
In the next section, we see that this difficulty is resolved for two meron configuration.

\section{Two merons and magnetic monopole loop}

\subsection{Meron pair solution}

The meron configuration is given by
\begin{align}
 g\mathbf{A}^{\rm M}_\mu(x) 
= g\mathbf{A}^A_\mu(x) \frac{\sigma_A}{2}  
= \frac{\sigma_A}{2} \eta^A_{\mu\nu} \frac{x_\nu}{x^2}  
 .
 \label{1-meron}
\end{align} 
We define the topological charge $Q_P$ using the topological charge density $D_P(x)$: 
\begin{align} 
 Q_P=\int d^4x D_P(x) , \quad 
 D_P(x) := \frac{1}{16\pi^2} {\rm tr}(\mathbf{F}_{\mu\nu}*\mathbf{F}_{\mu\nu}) 
  .
 \label{topological-density}
\end{align} 
The meron (\ref{1-meron}) has one half unit of topological charge $Q_P=1/2$ concentrated at the origin:
\begin{align} 
 D_P(x) = \frac12 \delta^4(x)   
 .
 \label{topological-density-1}
\end{align} 
This single meron solution
\begin{align}
 g\mathbf{A}_\mu^{\rm M}(x) = g\mathbf{A}^A_\mu(x) \frac{\sigma_A}{2} = S_{\mu\nu} \frac{x_\nu}{x^2} , \quad
 S_{\mu\nu} := - \frac{i}{4} (\bar{e}_\mu e_\nu - e_\nu \bar{e}_\mu) = \eta^A_{\mu\nu} \frac{\sigma_A}{2}
 ,
\end{align}
can be rewritten in another form:
\begin{align}
 g\mathbf{A}_\mu^{\rm M}(x) 
 = \frac12 i U(x) \partial_\mu U^{-1}(x) 
 ,
\end{align}
where
\footnote{
These relations are easily checked by using the formulae:
\begin{align}
 \bar{e}_\mu e_\nu = \delta_{\mu\nu} + i \eta^A_{\mu\nu} \sigma_A, \quad 
 e_\mu \bar{e}_\nu = \delta_{\mu\nu} + i \bar{\eta}^A_{\mu\nu} \sigma_A  
 .
\end{align}
}
\begin{align}
 U(x) = \frac{\bar{e}_\alpha x_\alpha}{\sqrt{x^2}}, \quad U^{-1}(x) = \frac{e_\alpha x_\alpha}{\sqrt{x^2}}  
 .
\end{align}
While, the single anti-meron solution
\begin{align}
 g\mathbf{A}_\mu^{\rm \bar{M}}(x) = g\mathbf{A}^A_\mu(x) \frac{\sigma_A}{2} = \bar{S}_{\mu\nu} \frac{x_\nu}{x^2} , \quad
 \bar{S}_{\mu\nu} := - \frac{i}{4} (e_\mu \bar{e}_\nu - \bar{e}_\nu e_\mu) = \bar{\eta}^A_{\mu\nu} \frac{\sigma_A}{2}
 ,
\end{align}
can be written in the form:
\begin{align}
 g\mathbf{A}_\mu^{\rm \bar{M}}(x) 
 = \frac12 i U^{-1}(x) \partial_\mu U(x) 
 .
\end{align}
Note that the meron and antimeron configurations are not of the pure gauge form, which has an important implications to confinement.

The single meron and anti-meron solutions given in the above are singular both at the origin $x^2=0$ and at infinity $x^2=\infty$.  
Using the conformal symmetry of the classical Yang-Mills action, it can be shown that in addition to a meron at the origin, there is a second meron at infinity with another half unit of topological charge.
In fact, the conformal invariance of Yang-Mills theory allows us to displace (map) those singularities to arbitrary points which we define to be the origin  and $d_\mu \in \mathbb{R}^4$.   Explicitly, the conformal transformation
\footnote{
This conformal transformation is obtained by combining
(a) a translation 
$
 x_\mu \rightarrow x_\mu +a_\mu
$,
(b) an inversion
$
 x_\mu \rightarrow - x_\mu/x^2
$,
(c) a dilatation (scale transformation)
$
 x_\mu \rightarrow -2a^2 x_\mu  
$, and
(d) a translation
$
 x_\mu \rightarrow x_\mu -a_\mu
$.
The transformation is constructed so that the origin $x=0$ is transformed to $a$, while $x=\infty$ to $-a$. In addition, $x=a$ is transformed to $0$.
}
\begin{align}
 x_\mu \rightarrow 
z_\mu =  2a^2 \frac{(x+a)_\mu}{(x+a)^2} - a_\mu   
 ,
 \label{conformal-transformation}
\end{align}
yields the new solutions
\begin{align}
 g\mathbf{A}_\mu^{\rm M}(x) \rightarrow \frac12 i U(z) \partial_\mu U^{-1}(z) := g\mathbf{A}_\mu^{\rm M\bar{M}}(x)   
 ,
 \nonumber\\
 g\mathbf{A}_\mu^{\rm \bar{M}}(x) \rightarrow  
 \frac12 i U^{-1}(z) \partial_\mu U(z) := g\mathbf{A}_\mu^{\rm \bar{M}M}(x) 
 ,
\end{align}
where
\footnote{
This is obtained from the transformation law:
$
 g\mathbf{A}_\mu^{\rm M}(x) \rightarrow \partial_\mu z_\nu   g\mathbf{A}_\nu^{\rm M}(z)
$.
} 
  $\partial_\mu:=\partial/\partial x_\mu \not= \partial/\partial z_\mu$. 
It is shown \cite{AFF77} that $g\mathbf{A}_\mu^{\rm M\bar{M}}$ corresponds to a meron located at $x=-a$ and an antimeron at $x=a$.  Conversely, $g\mathbf{A}_\mu^{\rm \bar{M}M}$ has a meron at $x=a$ and an antimeron at $x=-a$. 
These $g\mathbf{A}_\mu^{\rm M\bar{M}}$ ($g\mathbf{A}_\mu^{\rm \bar{M}M}$) are meron--antimeron (antimeron--meron) solutions. 
The meron-antimeron solution has the explicit expression for $a=(0,0,0,T)$:
\begin{align}
 g\mathbf{A}_\mu^{\rm M\bar{M}}(x) 
=&  \begin{cases}
   \frac{2T}{\tau^2}  x_4 \sigma_\ell x_\ell & (\mu=4) \cr
   \frac{2T}{\tau^2} [\epsilon_{jk\ell} T x_k \sigma_\ell + \frac12 (T^2-x^2) \sigma_j + x_j \sigma_\ell x_\ell ] & (\mu=j) 
   \end{cases}
 ,
 \nonumber\\
& \tau^2 = (T^2+x^2-2T x_4)(T^2+x^2+2T x_4) = (T^2+x^2)^2-4T^2x_4^2 
 .
\end{align}

It is also shown \cite{AFF77}  that the meron--meron (antimeron--antimeron) solution is given by performing a singular gauge transformation $U(y_{+})$ which changes the antimeron (meron) at $x=-a$ into a meron (antimeron) at the same point, leading from an $\rm M\bar{M}$ ($\rm \bar{M}M$) to an $\rm MM$ ($\rm \bar{M}\bar{M}$) one where
$y_{\pm}:=x\pm a$.
In fact, the singular gauge transformation
\begin{align}
  g\mathbf{A}_\mu^{\rm M\bar{M}}(x) \rightarrow U^{-1}(y_{+})g\mathbf{A}_\mu^{\rm M\bar{M}}(x)U(y_{+})+iU^{-1}(y_{+}) \partial_\mu U(y_{+}) := g\mathbf{A}_\mu^{\rm MM}(x)
 ,
 \label{gauge-transformation}
\end{align}
leads to the dimeron solution
\begin{align}
   g\mathbf{A}_\mu^{\rm MM}(x) = - S_{\mu\nu} \left[ \frac{y_{+}^\nu}{y_{+}^2} + \frac{y_{-}^\nu}{y_{-}^2} \right] 
= - \frac{\sigma_A}{2} \left[ \eta^A_{\mu\nu} \frac{(x+a)_\nu}{(x+a)^2} + \eta^A_{\mu\nu}  \frac{(x-a)_\nu}{(x-a)^2} \right]
 .
 \label{2-meron}
\end{align}
The antidimeron solution $g\mathbf{A}_\mu^{\rm \bar{M}\bar{M}}$ is obtained in the similar way. 

The gauge field $g\mathbf{A}_\mu^{\rm MM}(x)$ for a meron pair has infinite action density at $x=\{ 0, d \}$ and the logarithmic singularity of the action integral comes from the delta function concentration of topological charge:
\footnote{
Similar to instantons, a meron pair can be expressed in singular gauge by performing a large gauge transformation about the midpoint of the pair, resulting in a gauge field that falls off faster at large distance $\mathbf{A} \sim x^{-3}$.
} 
\begin{align}
 D_P(x)  = \frac12 \delta^4(x+a) + \frac12 \delta^4(x-a)  
 .
 \label{topological-density-3}
\end{align}

\subsection{Smeared meron pair}

\begin{figure}[ptb]
\begin{center}
\includegraphics[height=4.0cm,clip]{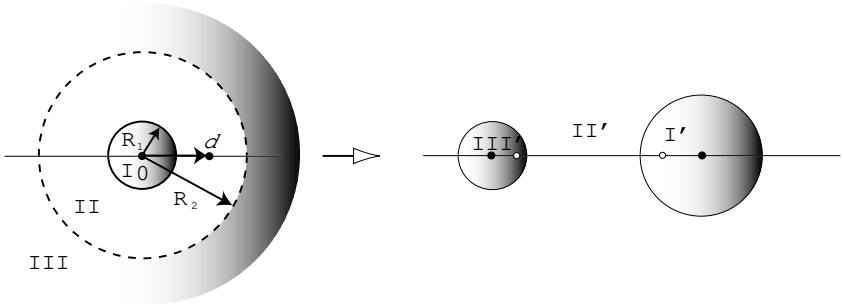}
\end{center} 
 \caption[]{
The concentric sphere geometry for a smeared meron (left panel) is transformed to the smeared two meron configuration (right panel) by the conformal transformation including the inversion about the point $d$. 
}
 \label{fig:smeared-meron}
\end{figure}

In order to eliminate the singularity in a meron pair configuration, we introduce an  Ansatz of  finite action by replacing the meron pair configuration (\ref{2-meron}) by a  smeared configurations  following Callan, Dashen and Gross \cite{CDG78} (See the left panel of Fig.~\ref{fig:smeared-meron}):
\begin{align}
 \mathbf{A}^{\rm sMM}_\mu(x) = \frac{\sigma_A}{2}  \eta^A_{\mu\nu} x_\nu  \times 
\begin{cases}
 \frac{2}{x^2+R_1^2} & $I$:\sqrt{x^2}<R_1 \cr
 \frac{1}{x^2} & $II$: R_1<\sqrt{x^2}<R_2 \cr
 \frac{2}{x^2+R_2^2} & $III$: \sqrt{x^2}>R_2  \cr
\end{cases} 
 ,
 \label{smeared-2-meron}
\end{align}
where  in the region (II) the field is identical to the meron field, while at the inner (outer) radius $R_1(R_2)$ it joins smoothly onto a standard instanton field.  Here the radii $R_1$ and $R_2$ of the inner sphere and the outer sphere  are arbitrary.

The topological charge $Q_P$ is spread out around the origin (I) and infinity (III): The scale size is chosen such that the net topological charge inside I (outside II)  is one-half unit, which agrees with the topological charge carried by each meron. 
This field (\ref{smeared-2-meron}) satisfies the equation of motion everywhere except on the two spheres.%
\footnote{
Although this patching of instanton caps is continuous, the derivatives are not, and therefore the equation of motion are violated at the boundaries of the regions, $\partial \rm I=\partial \rm II$ and $\partial \rm II=\partial \rm III$. 
}
  In fact, it is the solution of the equation of motion under the constraint that there be one-half unit of topological charge both in the inner and outer spheres, i.e., $Q_P^{\rm I}=1/2=Q_P^{\rm III}$. 
In other words, the singular meron fields for I and III are replaced by instanton caps, each containing topological charge 1/2 to agree with (\ref{smeared-2-meron}).

The Yang-Mills action of the new configuration is calculated to be
\begin{align}
 S^{\rm sMM}_{\rm YM} = \frac{8\pi^2}{g^2} +  \frac{3\pi^2}{g^2} \ln \frac{R_2}{R_1} 
 ,
\end{align}
where the first constant term comes from the two half-instantons in (I) and (III) 
\footnote{
There is no angular dependence in this patching, and so the conformal symmetry of the meron pair is retained.  For example, under a dilatation $x_\mu \rightarrow \lambda z_\mu$, both $R_1$ and $R_2$ get multiplied by $1/\lambda$, but the ratio and hence the action remain invariant. 
}
and the second logarithmic term comes from the pure meron region (II) in between.  Furthermore, if we let $|R_1-R_2| \downarrow 0$, this configuration becomes standard instanton. This is the one-instanton limit  ($R_2/R_1 \downarrow 1$). 
One meron limit is obtained by $R_2 \uparrow \infty$ or $R_1 \downarrow 0$ ($R_2/R_1 \uparrow \infty$). 

We perform the conformal transformation of the configuration about some point $d$ in the region II between $R_1$ and $R_2$:
\begin{align}
 x_\mu \rightarrow d_\mu + \rho^2 \frac{(x-d)_\mu}{(x-d)^2}
 ,
 \label{conformatl-transformation}
\end{align}
with $\rho$ an arbitrary scale factor. 
Because of conformal invariance, this produces an another acceptable solution of the equation of motion.  
The geometry before and after the  conformal transformation  is described in  Fig.~\ref{fig:smeared-meron}.
The conformal transformation maps a sphere into another sphere.
Therefore, the regions I and III, i.e., inner and outer spheres are transformed to two spheres, i.e., regions I' and III' with center coordinates $x_{\rm I'}$, $x_{\rm III'}$ and the scale sizes $R_1'$, $R_2'$, and  the field in region I' and III' is an instanton, since the conformal transformation of an instanton is again an instanton. 
Region II is transformed to region II' and the field in II' is given by
\begin{align}
 \mathbf{A}^{\rm II'}_\mu(x) = \frac{\sigma_A}{2} \left[ \eta^A_{\mu\nu} \frac{(x-x_{\rm I'})_\nu}{(x-x_{\rm I'})^2} + \eta^A_{\mu\nu}  \frac{(x-x_{\rm III'})_\nu}{(x-x_{\rm III'})^2} \right]
 ,
 \label{smeared-2-meron-b}
\end{align}
where  
\begin{align}
 x_{\rm I'} = \frac{R_2d}{R_2-R_1} > d ,
 \quad
 x_{\rm III'} = -\frac{R_1d}{R_2-R_1}<0
 .
\end{align}
The corresponding field strength $\mathbf{F}^{\rm II'}_{\mu\nu}$ falls at infinity as $|x|^{-4}$, leading to a convergent action integral.  
Since the topological charge $Q_p$ is conformal invariant, after transformation we have two spherical regions I', III' of net topological charge one-half surrounded by an infinite region II' of zero topological charge density $D_P(x)=0$.
Therefore, the transformed configuration is a smeared version of two merons at position $x_{\rm I'}$ and $x_{\rm III'}$.

\begin{figure}
\begin{center}
\includegraphics[height=3.0cm,clip]{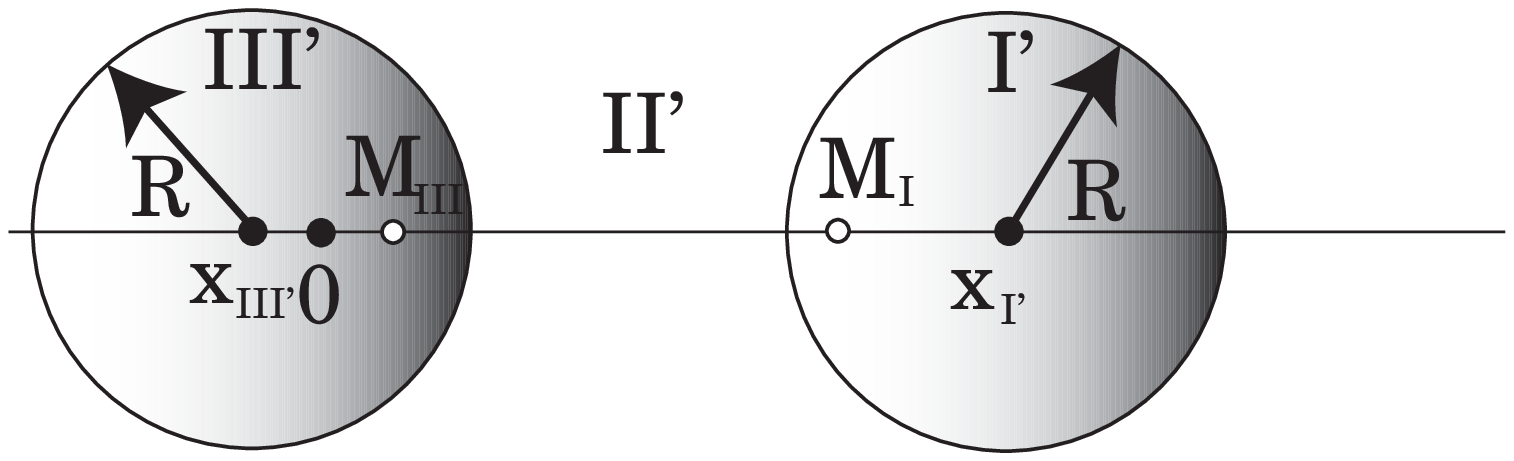}
\end{center} 
 \caption[]{
 Meron pair separated by $d=\sqrt{R_1 R_2}$ regulated with instanton caps. 
The smeared two meron configuration is obtained by the conformal transformation where  $d$ is the scale parameter of the inversion.
The centers of the sphere are 
$
 x_{\rm I'} = \frac{R_2d}{R_2-R_1} ,
$
and
$
 x_{\rm III'} = -\frac{R_1d}{R_2-R_1}
$.
The original positions of the two merons are not the centers of the sphere, nor are they the positions of maximum action density, which occurs with the spheres at 
$
 (M_{\rm I})_\mu = \frac{R_1^2}{R_1^2+d^2} d_\mu,
$
$ 
 (M_{\rm II})_\mu = \frac{R_2^2}{R_2^2+d^2} d_\mu, 
$
with 
$
 S_{\rm max}=\frac{48}{g^2}\frac{(R_1+R_2)^4}{d^8}
$.
The radius of the sphere is 
$
R = \frac{R_1 R_2}{R_2-R_1}
$. 
}
 \label{fig:smeared-meron2}
\end{figure}

The smoothed meron configuration may be thought of as describing various stages in a sequence of deformations of the instanton, leading from the instanton at one extreme to two widely separated smeared merons at the other. 
In a sense the meron is to be regarded as a constituent of the instanton.
This is realized by holding $R_1$ fixed and increasing $R_2$ from $R_1$ to infinity $\infty$.  For definiteness, we choose $\rho=d:=\sqrt{R_1 R_2}$, see Fig.~\ref{fig:smeared-meron2}.
With these choices, the configuration is two half instantons of scale size $R_1$ and separation $d=$ between the centers of the instanton configuration.
The action is 
\begin{align}
 S^{\rm sMM} = \frac{8\pi^2}{g^2} +  \frac{6\pi^2}{g^2} \ln \frac{d}{R_1} 
 .
\end{align}
As $R_2 \rightarrow R_1$, the regions I' and III' grow without limit in radius and move toward each other, while the centers of the instanton configurations approach each other and the region II' vanishes. 
In the limit the configuration is just given an instanton of scale size $\rho=d=R_1$ split in half through a center.

\subsection{Magnetic monopole loop joining the smeared meron pair}

Now we consider the solution of RDE for a smeared (regularized) meron pair configuration based on the finite action Ansatz.  For this purpose, we estimate the value of $\lambda(x)$ in each region. 
For a set of $(J,L)$, $\lambda(x)$ is calculated in each region as follows, see the next Table.

\begin{center}
\begin{tabular}{l|l|l|l|l|l|l} \hline\hline
J & L & S & degeneracy & I:$0<\sqrt{x^2}<R_1$ & II:$R_1<\sqrt{x^2}<R_2$ & III:$\sqrt{x^2}>R_2$ \\  
  &   &   &   & $\kappa=1, s=R_1$ & $\kappa=1/2, s=0$ & $\kappa=1, s=R_2$ \\
\hline\hline
1 & 0 & 1     & 3  & $\frac{8x^2}{(x^2+R_1^2)^2}$ & $\frac{2}{x^2}$ & $\frac{8x^2}{(x^2+R_2^2)^2}$ \\ \hline
1/2 & 1/2 & 1 & 4  & $\frac{2}{x^2-(\hat{b} \cdot x)^2}-\frac{8R_a^2}{(x^2+R_a^2)^2}$ & $\frac{2}{x^2-(\hat{b} \cdot x)^2}-\frac{2}{x^2}$ & $\frac{2}{x^2-(\hat{b} \cdot x)^2}-\frac{8R_a^2}{(x^2+R_2^2)^2}$ \\  
\hline
0 & 1 & 1     & 3  & $\frac{8x^2}{(x^2+R_1^2)^2}$ & $\frac{2}{x^2}$ & $\frac{8x^2}{(x^2+R_2^2)^2}$ \\ 
\hline
\end{tabular}
\label{table:eigenvalue2}
\end{center}

For $(J,L)=(1,0)$, 
\begin{align}
 \lambda(x) = V(x)  
=   2x^2 f^2(x)  
= \frac{8\kappa^2 x^2}{(x^2+s^2)^2}
= \begin{cases}
  \frac{8x^2}{(x^2+R_a^2)^2} &  $I$, $III$ \cr
  \frac{2}{x^2} & $II$ 
  \end{cases}
 .
\end{align}

For $(J,L)=(1/2,1/2)$, 
\begin{align}
 \lambda(x) 
= \frac{2}{x^2-(\hat{b} \cdot x)^2} + \frac{8\kappa^2 x^2}{(x^2+s^2)^2} - \frac{16\kappa}{x^2+s^2}
= \begin{cases}
  \frac{2}{x^2-(\hat{b} \cdot x)^2}-\frac{8R_a^2}{(x^2+R_a^2)^2} &  $I$, $III$ \cr
  \frac{2}{x^2-(\hat{b} \cdot x)^2}-\frac{2}{x^2} & $II$ 
  \end{cases}
 .
\end{align}

For $(J,L)=(0,1)$, 
\begin{align}
 \lambda(x)  
=\frac{8}{x^2} + \frac{8\kappa^2 x^2}{(x^2+s^2)^2} - \frac{8\kappa}{x^2+s^2}
= \begin{cases}
  \frac{8R_a^2}{x^2(x^2+R_a^2)^2} &  $I$, $III$ \cr
  \frac{2}{x^2} & $II$ 
  \end{cases}
 .
\end{align}

Comparing the above results, we find,  in order to make the  spacetime integral of  $\lambda(x)$ in each region as small as possible, that $(J,L)=(1,0)$ is selected for small $x$, i.e., the region I, $(J,L)=(0,1)$ is for large $x$, i.e., the region III, while in the intermediate region II, $(J,L)=(1/2,1/2)$ can give the smallest value of $\lambda(x)$. 
The result is summarized as
\begin{align}
 \lambda(x) 
= \begin{cases}
  \frac{8x^2}{(x^2+R_1^2)^2}  &  $I$: 0<\sqrt{x^2}<R_1; (J,L)=(1,0), \ \mathbf{n}_A(x)=Y^A_{(1,0)}=\text{const.} \cr
  \frac{2(\hat{b} \cdot x)^2}{x^2[x^2-(\hat{b} \cdot x)^2]} & $II$: R_1<\sqrt{x^2}<R_2; (J,L)=(\frac12,\frac12), \ \mathbf{n}_A(x) \simeq Y^A_{(1/2,1/2)}=\text{hedgehog} \cr
   \frac{8R_2^2}{x^2(x^2+R_2^2)^2} & $III$: R_2<\sqrt{x^2} ; (J,L)=(0,1), \ \mathbf{n}_A(x)=Y^A_{(0,1)}(x)=\text{Hopf}
  \end{cases}
 .
 \label{two-meron-lambda}
\end{align}

\begin{figure}[ptb]
\begin{center}
\includegraphics[width=12.5cm]{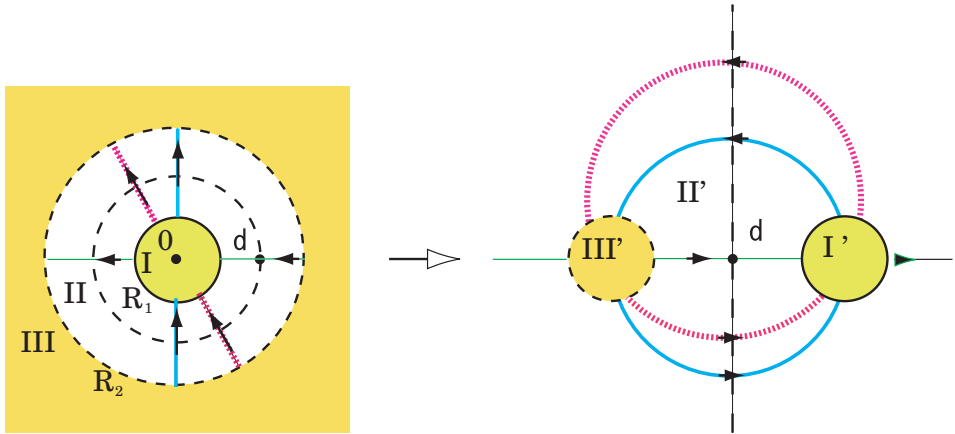}
\end{center} 
 \caption[]{
A magnetic monopole line in II connecting I and III along the direction of $\hat{b}_\mu$ in a smeared meron (left panel) is transformed to a circular magnetic monopole loop in II' connecting I' and III' connecting two merons. 
The magnetic monopole world line going through the center of inversion $d$ in (II) (left panel) is inverted to the  straight line connecting two merons (right panel), which is to be understood as the limit of the circle with infinite radius.  
Here the conformal transformation  maps a sphere to another sphere and preserves the angle between two vectors. 
The angle between two magnetic monopole lines in the left panel is preserved after the transformation in the right panel.  This is also the case for the angles between magnetic lines and concentric spheres in II.
}
 \label{fig:meron-monopole}
\end{figure}

As we have already shown in the previous section, the  magnetic current exists only for $(J,L)=(1/2,1/2)$.  Therefore, in the smeared meron pair configuration, the magnetic current flows only in the region II, while there is no magnetic current in regions I and III.
See the left panel of Fig.~\ref{fig:meron-monopole}. 
For the magnetic current parallel to $d_\mu$, the transformed magnetic current comes from the infinity, goes through two merons and  goes away to infinity, constituting the   straight line, see the right panel of Fig.~\ref{fig:meron-monopole}.
For the magnetic current orthogonal to   $d_\mu$ flowing from $\partial \rm I$ ($\partial \rm III$) to $\partial \rm III$ ($\partial \rm I$), the transformed magnetic current draws a piece of a circle beginning at $\partial \rm I'$ ($\partial \rm III'$) and ending at $\partial \rm III'$ ($\partial \rm I'$).  
Every magnetic current flowing in II is transformed to a circular magnetic monopole loop connecting $\rm I'$ and $\rm III'$.
See Fig.~\ref{fig:meron-monopole}. 
This is easily understood by considering intersections between magnetic lines and concentric spheres in II  from a fact that the conformal transformation  maps a sphere to another sphere and preserves the angle between two vectors. 

Note that $\lambda(x)$ obtained in (\ref{two-meron-lambda}) is always finite. 
In addition, due to the rapid decrease of $\lambda(x)$, $\lambda(x) \sim O(x^{-6})$  in region III  
and the asymptotic behavior $\lambda(x) \sim O(x^{2})$  in region I, the reduction functional becomes finite:
\begin{align}
 F_{\rm rc} = \int_{\mathbb{R}^4} d^4x \lambda(x) < \infty
  ,
\end{align}
as far as $R_1, R_2 > 0$.
Therefore, this is an allowed solution. 
Thus we have obtained  circular magnetic monopole loops with a non-zero radius $r \ge d/2$  joining a meron pair separated by a distance $d$.

It should be remarked that the reduction functional (\ref{Rc}) is conformal invariant. 
Therefore the color field in the region (II') for the meron pair is given by the conformal transformation  (\ref{conformal-transformation}) and a subsequent singular gauge transformation $U(y_{+})$ (\ref{gauge-transformation}):
\begin{align}
\bar{\mathbf{n}}(x)_{\rm II'} 
= \frac{2a^2}{(x+a)^2}
\hat{b}_\nu \eta^A_{\mu\nu} z_\mu U^{-1}(y_{+})  \sigma_A U(y_{+})/\sqrt{ z^2 -(\hat{b} \cdot z)^2}
 ,
\end{align}
where $z$ is given by (\ref{conformal-transformation}) and $y_+$ is the same as that in (\ref{gauge-transformation}). 
The color field in the region (III') for the meron pair  can be obtained by applying the conformal transformation (\ref{conformal-transformation}) and the gauge transformation (\ref{gauge-transformation}) to the standard Hopf map.
The color field  in the region (I') for the meron-meron  is trivial. 

The location of the magnetic monopole is dictated by the simultaneous zeros of $\hat{b}_\nu \eta^A_{\mu\nu} z_\mu$ for $A=1,2,3$:
\begin{align}
 0 
 = \hat{b}_\nu \eta^A_{\mu\nu} [2a^2(x_\mu+a_\mu)-(x+a)^2 a_\mu ]
 \quad (A=1,2,3)
 ,
 \label{zeros}
\end{align}
since the gauge transformation $U(y_{+})$ does not change the zeros. 
Without loss of generality, we can fix the direction of connecting two merons as
$a_\mu:=d_\mu/2=\delta_{\mu4}T$. 
For $a_\mu =\delta_{\mu4}T$,  
$
\hat{b}_\nu \eta^A_{\mu\nu}a_\mu
= \epsilon_{Ajk}\hat{b}_k a_j +a_A \hat{b}_4-\hat{b}_A a_4 =-\hat{b}_AT
$ 
and (\ref{zeros}) reads
\begin{align}
 \hat{b}_A x^2  + 2T \hat{b}_k \epsilon_{Ajk}   x_j  + 2T\hat{b}_4  x_A  - \hat{b}_A T^2 = 0 
\quad (A=1,2,3)
 .
 \label{zeros-2}
\end{align}

It is instructive to see two special cases. 
If $\hat{b}_\mu$ is parallel to $a_\mu$, i.e., $\hat{b}_\mu=\delta_{\mu 4}$ (or $\mathbf{\hat{b}}=\mathbf{0}$), we find from (\ref{zeros-2}) that the simultaneous zeros are given by $x_A=0$ ($A=1,2,3$), i.e., the magnetic current is located on the $x_4$ axis which is parallel to $a_\mu$.   The magnetic monopole curret denotes a straight line going through two merons at $(\mathbf{0}, \pm T)$. See a horizontal line in the right panel of Fig.~\ref{fig:meron-monopole}.
This straight line can be identified with the maximal circle with infinite radius in the general case discussed below. 

If $\hat{b}_\mu$ is perpendicular to $a_\mu$ (or $\hat{b}_\mu=\delta_{\mu \ell} \hat{b}_\ell$, $\ell=1,2,3$), i.e., $ \hat{b}_4=0$,  
the simultaneous zeros are obtained on a circle 
\begin{align}
 x_\ell^2 + x_4^2 = T^2  
 .
\end{align} 
In this case, the circular magnetic monopole loop has its center at the origin $0$ in $z$ space and the radius $T=\sqrt{a^2}$ joining two merons at $(\mathbf{0}, \pm T)$ on the plane spanned by $a_\mu$ and $\hat{b}_\ell$. 
See a minimal circle in the right panel of  Fig.~\ref{fig:meron-monopole}.  

In general, it is not difficult to show that the simultaneous zeros are given by   
\begin{align}
 \mathbf{x} \times \mathbf{\hat{b}} = \mathbf{0}
 \quad \& \quad 
 \left( \mathbf{x} + T \frac{\hat{b}_4}{|\mathbf{\hat{b}}|} \frac{\mathbf{\hat{b}}}{|\mathbf{\hat{b}}|} \right)^2 + x_4^2 = T^2 \left( 1 +    \frac{\hat{b}_4^2}{|\mathbf{\hat{b}}|^2}  \right)
 ,
\end{align}
where $\mathbf{\hat{b}}$ is the three-dimensional part of unit four vector $\hat{b}_\mu$ ($\hat{b}_\mu \hat{b}_\mu=\hat{b}_4^2+|\mathbf{\hat{b}}|^2= 1)$.
These equations express circular magnetic monopole loops  joining two merons at $\pm a_\mu$ on the plane  specified by $a_\mu$ and $\mathbf{\hat{b}}$ where a  circle has the center at 
$
\mathbf{x}=-T \frac{\hat{b}_4}{|\mathbf{\hat{b}}|} \frac{\mathbf{\hat{b}}}{|\mathbf{\hat{b}}|}
=-\sqrt{a^2} \frac{\hat{b}_4}{|\mathbf{\hat{b}}|} \frac{\mathbf{\hat{b}}}{|\mathbf{\hat{b}}|}
$ and $x_4=0$ with the radius $T\sqrt{1 +     \frac{\hat{b}_4^2}{|\mathbf{\hat{b}}|^2}}=\frac{\sqrt{a^2}}{|\mathbf{\hat{b}}|}( \ge T)$. 
See a larger circle in the right panel of Fig.~\ref{fig:meron-monopole}.
The horizontal straight line can be identified with the limit of  infinite radius of the circle.


Finally, we can reproduce the one-instanton case by considering the one-instanton limit  $R_2 \rightarrow R_1$ of the meron pair.
In the one-instanton limit,  the region II' vanishes and the magnetic monopole loop disappears.  This reproduces the previous result \cite{BOT97}  that a circular magnetic monopole loop is shrank to the center of an instanton in one-instanton background field.  
Thus the instanton can not be the quark confiner which is consistent with the dual superconductivity picture for quark confinement where the magnetic monopole loop must be  the dominant configuration responsible for confinement.

\section{Conclusion and discussion}

In this paper, we have examined in an analytical way whether circular loops of magnetic monopole \cite{KMS06,Kondo08} exist or not for a given background of Yang-Mills field in the four-dimensional Euclidean SU(2) Yang-Mills theory. As  Yang-Mills background fields, we have examined  some known solutions of the Yang-Mills field equation of motion, i.e., one-instanton, one-meron and two-merons. 
The analysis has been performed using the recently developed reformulation of Yang-Mills theory \cite{KMS06,KSM08}.

Consequently, we have obtained a new analytical result that there exist circular magnetic monopole loops  supported by a pair of merons smeared  (ultraviolet regularized) in the sense of Callan, Dashen and Gross \cite{CDG78}, although the corresponding numerical solution has already been found by Montero and Negele \cite{MN02} on a lattice.
Moreover, we have reproduced some of previous results in the same reformulation, which have been obtained in specific (partial) gauge fixing procedures called MAG,  LAG and MCG in which topological objects such as Abelian magnetic monopoles and center vortices are regarded as gauge-fixing defects. 
We have obtained the corresponding gauge-invariant results: 
(1) One  instanton configuration can not support a (gauge-invariant) magnetic monopole loop.
(2) One meron configuration induces a (gauge-invariant)  magnetic monopole current along a straight line going through the meron. 
However, neither one-instanton nor one-meron supports circular magnetic monopole loops.

As we have shown in this paper, a meron pair is a first topological object which is found to be consistent with the dual superconductor picture of quark confinement. 
Therefore, the meron pair configurations are candidates for field configurations to be responsible for deriving the area law of the Wilson loop average. 
The detailed analysis will be given in a subsequent paper \cite{Kondo08c} which forms also a relationship  with the recent  papers \cite{Kondo08} and \cite{Kondo08b}.


\section*{Acknowledgments}

The authors would like to thank Takeharu Murakami for helpful discussions in the early stage of this work. 
This work is financially supported by Grant-in-Aid for Scientific Research (C) 18540251 from Japan Society for the Promotion of Science
(JSPS).

\appendix
\section{Deriving the reduction differential equation}\label{appendix:rc}

The infinitesimal form of the enlarged gauge transformation for $\delta{\bm A}_\mu$ and $\delta\bm \phi$ is given by \cite{KMS06}
\begin{align}
\delta_\omega \mathbf{A}_\mu(x)=& D_\mu[\mathbf{A}]\bm\omega(x) ,
\quad
\delta_\theta \bm \phi(x) =g\bm \phi(x) \times\bm\theta_\perp(x)
\nonumber\\
& (\bm\omega \in SU(2),\bm\theta_\perp \in SU(2)/U(1))
\label{egt}
 ,
\end{align}
where the subscript $\perp$ denotes the components perpendicular to $\bm \phi$. 

We wish to minimize the functional
\begin{align}
  R[\mathbf{A},  \bm \phi]
  =    \int d^Dx
   \frac12 (D_\mu[\mathbf{A}]\bm \phi)
   \cdot (D_\mu[\mathbf{A}]\bm \phi)
 ,
\end{align}
with respect to the enlarged gauge transformation as 
\begin{align}
0 = \delta R[\mathbf{A},  \bm \phi]
 &= 
   \int d^Dx
   D_\mu[\mathbf{A}]\bm \phi
   \cdot\delta(D_\mu[\mathbf{A}]\bm \phi)
   \nonumber\\
 &= 
   \int d^Dx
   D_\mu[\mathbf{A}]\bm \phi
   \cdot
   (D_\mu[\mathbf{A}]\delta\bm \phi
    +g\delta\mathbf{A}_\mu\times\bm \phi)
   \nonumber\\
 &= 
   \int d^Dx \{
   D_\mu[\mathbf{A}]\bm \phi
   \cdot
   (D_\mu[\mathbf{A}](g\bm \phi\times\bm\theta_\perp)
    +g(D_\mu[\mathbf{A}]\bm\omega)\times\bm \phi \} 
   \nonumber\\
 &= 
   \int d^Dx \{
   D_\mu[\mathbf{A}]\bm \phi
   \cdot
   (D_\mu[\mathbf{A}](g\bm \phi\times\bm\theta_\perp)
    +g D_\mu[\mathbf{A}](\bm\omega \times\bm \phi) \} 
   \nonumber\\
 &=  g
   \int d^Dx
   D_\mu[\mathbf{A}]\bm \phi
   \cdot
   D_\mu[\mathbf{A}]\{\bm \phi \times(\bm\theta_\perp-\bm\omega)\}
   \nonumber\\
 &=  g
   \int d^Dx
   (D_\mu[\mathbf{A}]\bm \phi)
   \cdot
   D_\mu[\mathbf{A}]\{\bm \phi \times(\bm\theta_\perp-\bm\omega_\perp)\}
 .
\end{align}
The integration by parts yields  
\begin{align}
0 = \delta R[\mathbf{A},  \bm \phi]
 &=-  g
    \int d^Dx
    (D_\mu[\mathbf{A}]D_\mu[\mathbf{A}]\bm \phi)
   \cdot
   \{\bm \phi \times(\bm\theta_\perp-\bm\omega_\perp)\}
   \nonumber\\
 &=  g
   \int d^Dx
   (\bm\theta_\perp-\bm\omega_\perp)
   \cdot\left(\bm \phi\times D_\mu[\mathbf{A}]D_\mu[\mathbf{A}]\bm \phi \right)
 .
\end{align}

Therefore, this functional is invariant if $\bm\omega_\perp=\bm\theta_\perp$. 
Thus, if $\bm\theta_\perp \ne  \bm\omega_\perp$, the minimization of the functional is achieved by $\bm \phi$ and $\bm A$ satisfying the differential equation:
\begin{align}
    \bm \phi\times D_\mu[\mathbf{A}]D_\mu[\mathbf{A}]\bm \phi   = 0
 .
\end{align}
This equation gives two conditions, since it is perpendicular to $\bm \phi (x)$.

\section{Simplifying the covariant Laplacian}\label{appendix:RDE}

 From the definition of the covariant derivative, we have
\begin{align}
 & D_\mu[\mathbf{A}] D_\mu[\mathbf{A}] \bm{\phi}
\nonumber\\ 
 =& \partial_\mu (D_\mu[\mathbf{A}] \bm{\phi}) + g \mathbf{A}_\mu \times (D_\mu[\mathbf{A}] \bm{\phi})
\nonumber\\ 
 =& \partial_\mu (\partial_\mu \bm{\phi} + g \mathbf{A}_\mu \times \bm{\phi}) + g \mathbf{A}_\mu \times (\partial_\mu \bm{\phi} + g \mathbf{A}_\mu \times \bm{\phi})
\nonumber\\ 
 =& \partial_\mu \partial_\mu \bm{\phi} + g \partial_\mu \mathbf{A}_\mu \times \bm{\phi}  + 2g \mathbf{A}_\mu \times  \partial_\mu \bm{\phi} +  g \mathbf{A}_\mu \times (g \mathbf{A}_\mu \times \bm{\phi})
\nonumber\\ 
 =& \partial_\mu \partial_\mu \bm{\phi} + g \partial_\mu \mathbf{A}_\mu \times \bm{\phi}  + 2g \mathbf{A}_\mu \times  \partial_\mu \bm{\phi} +  (g \mathbf{A}_\mu \cdot \bm{\phi}) g \mathbf{A}_\mu - (g \mathbf{A}_\mu \cdot g \mathbf{A}_\mu)  \bm{\phi} 
 .
\end{align}
We adopt the Ansatz
\begin{align}
 g \mathbf{A}_\mu^A(x) = \eta^A_{\mu\nu} f_\nu(x) 
 = \eta^A_{\mu\nu} \partial_\nu \ln \Phi(x) 
 .
\end{align}
First, the Yang-Mills field satisfies the Lorentz condition, i.e., divergenceless:
\begin{align}
 \partial_\mu \mathbf{A}_\mu^A(x) 
 = g^{-1} \eta^A_{\mu\nu} \partial_\mu f_\nu(x) 
 = g^{-1} \eta^A_{\mu\nu} \partial_\mu \partial_\nu \ln \phi(x)
= 0 
 ,
\end{align}
where we have used $\eta^A_{\mu\nu}=-\eta^A_{\nu\mu}$.
Moreover, we find
\begin{align}
  (g \mathbf{A}_\mu \cdot \bm{\phi}) g \mathbf{A}_\mu
 =& (g \mathbf{A}_\mu^B \bm{\phi}^B) g \mathbf{A}_\mu
\nonumber\\ 
 =& \eta^B_{\mu\beta} f_\beta \bm{\phi}^B \eta^B_{\mu\alpha} f_\alpha
\nonumber\\ 
 =& (\delta_{AB}\delta_{\alpha\beta}+\epsilon_{ABC}\eta^C_{\alpha\beta})  f_\alpha f_\beta \bm{\phi}^B
\nonumber\\ 
 =&  f_\alpha f_\alpha \bm{\phi}^A
 ,
\end{align}
and
\begin{align}
(g \mathbf{A}_\mu \cdot g \mathbf{A}_\mu)  \bm{\phi} 
= \eta^B_{\mu\alpha} f_\alpha \eta^B_{\mu\beta} f_\beta   \bm{\phi}^A
= 3 f_\alpha f_\alpha \bm{\phi}^A
 ,
\end{align}
where we have used
$
\eta^A_{\mu\alpha} \eta^B_{\mu\beta}
= \delta_{AB}\delta_{\alpha\beta}+\epsilon_{ABC}\eta^C_{\alpha\beta}
$.
Finally, we have
\begin{align}
(2g \mathbf{A}_\mu \times  \partial_\mu \bm{\phi})_A 
= 2 \epsilon_{ACB} \mathbf{A}_\mu^C \partial_\mu \phi_B 
= 2 \epsilon_{ACB} \eta^C_{\mu\nu} f_\nu \partial_\mu \phi_B 
 .
\end{align}
Thus we arrive at
\begin{align}
  (-D_\mu[\mathbf{A}] D_\mu[\mathbf{A}] \bm{\phi})_A
=  -\partial_\mu \partial_\mu \phi_A  
+2 f_\alpha f_\alpha \phi_A 
 - 2 \epsilon_{ACB} \eta^C_{\mu\nu} f_\nu \partial_\mu \phi_B
 .
\end{align}


\end{document}